\documentclass{aa}  

\usepackage{graphicx}
\usepackage{subcaption}
\usepackage{orcidlink}
\usepackage{txfonts}
\usepackage{multirow}
\usepackage{relsize}

\usepackage{sidecap}
\begin{document}

   \title{New classification method for the dynamical state of galaxy clusters with a Gaussian mixture model}
  \titlerunning{GMM classification for the dynamical state of clusters} 
   \subtitle{}

   \author{Hyowon Kim\inst{1,2,3}\orcidlink{0000-0003-4032-8572}
          \and Marco Canducci\inst{4}\orcidlink{0000-0003-2264-9743}
          \and Rory Smith\inst{1,2}\orcidlink{0000-0001-5303-6830}
          \and Peter Tino\inst{4}
          \and Yara Jaffe\inst{1,2}\orcidlink{0000-0003-2150-1130}
          \and Ho Seong Hwang\inst{5,6}\orcidlink{0000-0003-3428-7612}
          \and Jihye Shin\inst{3,7}\orcidlink{0000-0001-5135-1693}
          \and Kyungwon Chun\inst{3}\orcidlink{0000-0001-9544-7021}
        } %\fnmsep
   \institute{Departamento de Fisica, Universidad Tecnica Federico Santa Maria, Avenida España 1680, Valpara$\acute{i}$so, Chile \label{inst1}
    \and Millenium Nucleus for Galaxies (MINGAL)\label{inst2}
    \and Korea Astronomy and Space Science Institute, Daejeon 34055, Republic of Korea \label{inst3}
    \and School of Computer Science, University of Birmingham, Edgbaston, Birmingham, B15 2TT, UK \label{inst4}
    \and Astronomy Program, Department of Physics and Astronomy, Seoul National University, 1 Gwanak-ro, Gwanak-gu, Seoul 08826, Republic of Korea\label{inst5}
    \and SNU Astronomy Research Center, Seoul National University, 1 Gwanak-ro, Gwanak-gu, Seoul 08826, Republic of Korea\label{inst6}
   \and  University of Science and Technology (UST), Gajeong-ro, Daejeon 34113, Republic of Korea\label{inst7}}
\authorrunning{Kim et al}

   \date{Received Septmber, 04, 2025; accepted March, 03, 2026}

% \abstract{}{}{}{}{} 
% 5 {} token are mandatory
 
  \abstract
   {Galaxy clusters are the largest gravitationally bound systems, and they continue their growth through mergers in a hierarchical $\Lambda$CDM Universe. Therefore, we can describe the merger stage of a cluster as the dynamical state of clusters. Previous studies have investigated this phenomenon, but several limitations remain, including reliance on dichotomous classifications, constraints on the number of indicators used, absence of reliability, and incompatibility of methods between observation and simulation studies.}
   {To overcome the limitations, we developed an enhanced and observation-applicable cluster dynamical state classification method using the Bayesian classifier with the class-conditional Gaussian mixture distribution model using the N-cluster Run simulation data.}
   {The Bayesian classifier was designed for two merger stages (merger and relaxed) as well as three merger stages (recent merger, ancient merger, and relaxed) to provide a more detailed interpretation of the merger processes. After the best classifier model was constructed, we applied it to the observation data to test its performance and usability.}
   {In the results, using a larger number of indicators yields better results, with their order of importance being: magnitude difference, center offset, sparsity, Kuiper V statistic, and mirror asymmetry. Additionally, our analyses show that a projected classifier (built on the 6D space, but evaluated on lower dimensional projections) consistently produces better outcomes than non-projected classifiers (i.e., classifiers built directly on the corresponding low dimensional spaces), which means limited observation data can be used to classify with enhanced performance. Furthermore, the new classification method outperforms our previous research. }
   {This new method can suggest a way of overcoming previous limitations and provides new insights by providing the reliability of dynamical state classification results. We expect this enhanced method and its findings can be used in observational studies to better understand the evolution of galaxy clusters and the mass assembly history of the Universe.}


   \maketitle
%
%-------------------------------------------------------------------

\section{Introduction}
In the hierarchical structure formation paradigm of the $\Lambda$CDM Universe, matter grows from the assembly of small components and consists of large structures through the merger of large systems \citep{2012ARA&A..50..353K}. Galaxy clusters are the largest gravitationally bound systems in the nearby Universe and continue to experience active mass assembly through mergers. Since mergers create conflict and disturbances among the components of the system, we can understand the process of cluster merger in terms of virialization (or relaxation), which reflects the dynamical state of clusters.

The distribution of galaxy clusters and their dynamical states serves as a direct indicator for estimating the evolution of matter in the Universe. Additionally, it plays a crucial role in understanding the evolution of galaxies in densely populated environments \citep[e.g.,][] {2015MNRAS.452.3030T,2020MNRAS.495..705Z,2025A&A...699A.313A}. Therefore, measuring the dynamical states of galaxy clusters with large survey data is essential for connecting various studies related to large-scale and small-scale cosmic evolution.

Numerous studies have explored the dynamical state of galaxy clusters; however, there are limitations in accurately describing the continuous transitions of dynamical states and in utilizing large photometric survey data. Due to the limited data available for clusters, criteria for defining dynamical states are often simplistic, classifying them as either relaxed or unrelaxed based on a single criterion. Alternatively, researchers need to combine different observational results across various wavelengths, which can be both costly and time-consuming \citep{2024ApJ...967...14C}. Recent studies have sought to address these limitations by integrating multiple indicators \citep{2020ApJ...904...36Z, 2020MNRAS.497.5485Y, 2021MNRAS.504.5383D,2022MNRAS.514.5890L,2022A&A...665A.117C, 2024MNRAS.532.1031H, 2024ApJ...967...14C}. These studies have provided valuable insights for how to improve dynamical state classification methods. However, it is often difficult to quantify the reliability of their classification results and sometimes the definition of the dynamical states was not clearly defined.

In our previous study \citep{2024ApJ...970..165K}, we aimed to develop more steps of criteria for the detailed separation of the dynamical states based on the merger stage of galaxy clusters. We attempted to describe these criteria by analyzing the success rate of separation along a one dimensional axis (1D), which is a rotated axis derived from a multidimensional indicator distribution, used to quantify reliability. This approach successfully classified clusters into recent, ancient, and relaxed merger stages, but it also has some limitations.

We were not able to consider more than four indicators simultaneously due to the constraints of mathematical calculations involved in the rotation matrix, the process required substantial computational time. We were limited to providing 1D recipes, as the method was too complicated for broader application, which lost some part of the information. 

To address these limitations, we are exploring a new approach that utilizes the Bayesian classifier with class-conditional Gaussian mixture model (GMM) through machine learning techniques to apply it to a large volume of optical survey data.

%layout of paper
This paper is organized as follows. Section \ref{sec:data} introduces our simulation and observational data, along with the merger stage sampling methods for the classification class. In Section \ref{sec:method}, we describe six dynamical indicators and the Bayesian classifier with the class-conditional GMM implemented using a machine learning approach. Section \ref{sec:result} presents the explanation of the modeling process, as well as the results of applying the classifier to the observational data. In Section \ref{sec:discussion}, we discuss the dependency of indicators on redshift and mass, compare our current findings with our previous study, and describe potential limitations. We conclude our study in Section \ref{sec:conclusion}.
The following cosmological parameters are assumed throughout this paper: $\Omega_{m} $= 0.3, $\Omega_{\Lambda}$= 0.7, $\Omega_{b}$= 0.047, and h = 0.684.

\section{Data}\label{sec:data}
\subsection{N-cluster Run simulation}
\begin{figure}
\centering
\includegraphics[scale=.75]{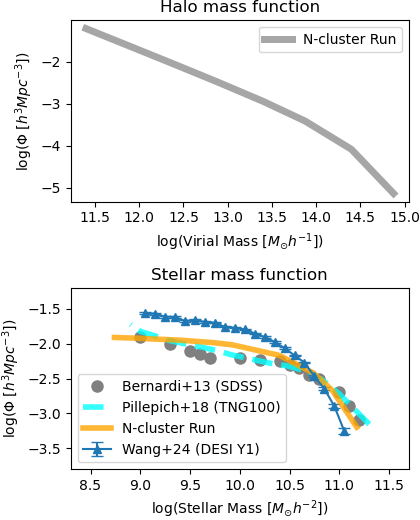}
\caption{Top: Halo mass function of N-cluster Run simulation. Bottom: Comparison with stellar mass function of the N-cluster Run simulation and other studies. The Orange solid and cyan dashed lines represent the simulation function from the N-cluster Run and the Illustris-TNG 100 simulation \citep{2018MNRAS.475..648P}. Gray circles and blue triangles are observational function from SDSS \citep{2013MNRAS.436..697B} and DESI Y1 data \citep{2024ApJ...971..119W}.}\label{fig:lf}
\end{figure}

Our goal in this work is to develop a practical observation-based method by utilizing realistic stellar component parameters of the brightest cluster galaxy (BCG). To achieve this, we employed simulation data from the N-cluster Run, consistent with previous research \citep{2024ApJ...970..165K}. The \texttt{N$\_$cluster run} simulation is a cosmological N-body dark matter-only simulation that consists of 64, each 120 Mpc $h^{-1}$ length cubic boxes. It has 169 snapshots with 100 megayear time resolution from z=200. Smallest particle mass is $1.07189\times10^9 M_{\sun}h^{-1}$. 
The \texttt{Gadget3} \citep{2005MNRAS.364.1105S} code is used for cosmological N-body/SPH simulations, and the 6D friends-of-friends (FoF) algorithm, \texttt{ROCKSTAR} halo finder \citep{2013ApJ...762..109B}, was used to define the halos of galaxies and clusters.

The N-cluster Run simulation is the dark matter-only simulation that uses the assumption of abundance matching to assign stellar masses to halo masses. Abundance matching, by definition, matches the simulated halo mass functions to observed stellar mass functions. By taking this approach, we ensure that LCDM simulations will reproduce the observed stellar mass functions over a wide range of redshifts. Our simulation results are comparable with the observations for the specific dynamical state indicators we used.

In this study, we expanded the sample size to three times larger than that of the earlier study, using 30 boxes containing a total of 1,845 galaxy clusters. The mass range of cluster size halos is from $2.68\times10^{13}M_\odot$ to $ 1.28\times10^{15} M_\odot$ and the galaxy stellar mass range is $2.3\times 10^6M_\odot$ to $1.59\times 10^{11}M_\odot$. Halo mass and stellar mass functions can be checked in Figure \ref{fig:lf}. Redshift is only considered from 0 to 0.5 (5 Gyr) to reduce the redshift dependency (See in section \ref{sec:dependency}). Further details of simulation can be found in these papers \citep{2021ApJ...912..149S, 2022AJ....164...95S, 2022ApJ...934...86S, 2022ApJ...925..103C, 2022ApJ...940....2J, 2022ApJ...935...71K, 2022ApJS..261...28Y, 2023ApJ...943..148C, 2023MNRAS.520.4517A, 2024MNRAS.527.9185D, 2024ApJ...969..142C, 2024ApJ...970..165K} and on the publically available data archive page \footnote{https://data.kasi.re.kr/vo/N$\_$cluster$\_$run}.

\subsection{Definition of the dynamical states for the classification class}
Merger stages for the dynamical states are categorized using the same approach outlined in the previous paper \citep{2024ApJ...970..165K}. We define the beginning of a merger as the moment when the infall halo crosses one virial radius from the main halo. This distance is calculated in 3D space (considering x, y, and z coordinates). The time since infall is measured in one-gigayear increments from the merger start epoch. To avoid confusion from multiple mergers, we only considered a single major merger (mass ratio greater than 1:5) within the redshift range of 0 < z < 0.5 as a disturbed sample. The relaxed state is defined as having a merger mass ratio of less than 1:10 within the same redshift range, which includes minor mergers. Minor mergers are not considered as a separate state in this study because of their weak signal. We find that minor mergers do not have a significant effect on the dynamical state indicators. 

We further divided the merger stages into recent and ancient mergers to provide a more detailed description of the dynamical state of clusters. The recent merger stage includes indicators that change at the start of the merger (one gigayear after the merger occurs), while the ancient merger stage encompasses the transitional phase of mergers (beginning three gigayears after the merger occurs). Table \ref{tab:mergertype} summarizes the dynamical states employed in this study.

In this work, we developed our methods using two dynamical state categories (merger and relaxed) and three dynamical states (recent merger, ancient merger, and relaxed) to compare our results with previous studies and offer a detailed classification of dynamical states. This information was used for training the classifier and testing the performance of the method.
\begin{table}
	\centering
	\caption{Summary of the dynamical states used in this study.}\label{tab:mergertype}
        \begin{tabular}{c|cc|c}
        \hline
        \multirow{2}{*}{Dynamical state } & \multicolumn{2}{c|}{   merger} & \multirow{2}{*}{relax} \\ \cline{2-3}
                          & \multicolumn{1}{c|}{recent} & ancient &   \\ \hline
      Time since infall   & \multicolumn{1}{c|}{1Gyr} & $\geq$3Gyr &    - \\ \hline
      Merger mass ratio   & \multicolumn{2}{c|}{$\geq$1:5}   & $\leq$1:10 \\ \hline
        \end{tabular}
        \tablefoot{Mass ratio and time were used as criteria.}
\end{table}

\subsection{Observation data}
To evaluate the application of our method using observational data, we utilized the Hectospec Cluster Survey catalog \citep[HeCS]{2013ApJ...767...15R}. This catalog comprises 211 clusters located in the northern hemisphere, with redshifts ranging from 0 to 0.3 and a mass range of \(9.5 \times 10^{13}\ M_\odot\) to \(6.13 \times 10^{14} M_\odot\). We chose total 135 clusters, 69 from CIRS \citep{2006AJ....132.1275R}, 25 from HeCS \citep[][HeCS]{2013ApJ...767...15R}, 8 from HeCS-red \citep{2018ApJ...862..172R}, 29 from HeCS-SZ \citep{2016ApJ...819...63R}, 1 from KYDISC \citep{2018ApJS..237...14O}, 1 from OmegaWINGS \citep{2017A&A...599A..81M}, and 2 clusters from the NASA/IPAC Extragalactic Database (NED1). We excluded some low statistics clusters from the original sample \citep{2023MNRAS.525.4685S}.

The average number of member galaxies per cluster is 100, with a range from 23 to 1,350. Due to observation limitations, membership of clusters tends to contain a higher number of bright red galaxies compared to blue, faint galaxies. Nevertheless, the data possesses adequate spectroscopic completeness (comp$\geq$0.5 in r-band$\geq$17.7) to allow for meaningful comparisons with simulation data. 

\begin{figure}
\centering
\includegraphics[scale=.75]{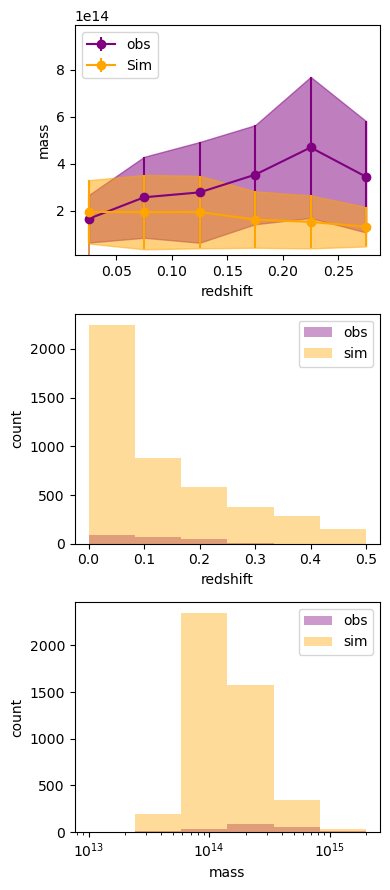}
\caption{Top: Mass versus redshift distribution of simulation data and observation data. Dots show median values, and the shaded area shows standard deviation values within redshift bins. Middle: Histogram of redshift for both observation and simulation data. Bottom: Histogram of Mass for both observation and simulation data.}\label{fig:mz_so}
\end{figure}

Figure \ref{fig:mz_so} shows the redshift and mass range of the observed and simulated clusters we used in this study. The redshift range of the simulation is much larger than that of the observation data and the sample size is significantly different. This can offer one proof that observation data can serve as a test sample of the simulation data-trained model.

\section{Methods}\label{sec:method}
In this section, we introduce six dynamical state indicators for galaxy clusters and explain their significance. We then describe the Bayesian classifier and class-conditional distribution with the GMM method, which takes into account the 6D covariances to model each dynamical state. In addition, we present the quantification method along with precision, recall, and accuracy metrics.

\subsection{Dynamical indicators}\label{sec:ind}
\begin{figure*}
\centering
\includegraphics[scale=.40]{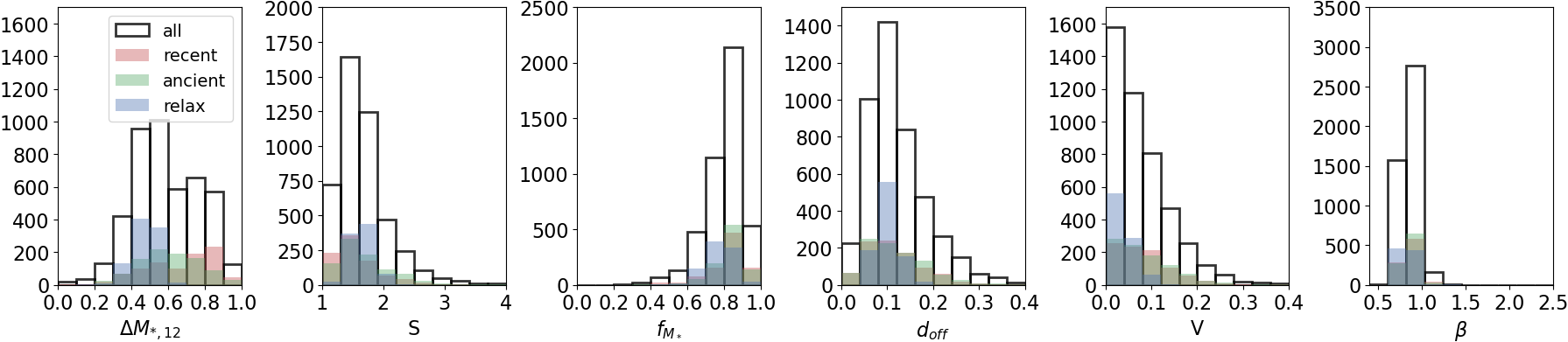}\\
\includegraphics[scale=.39]{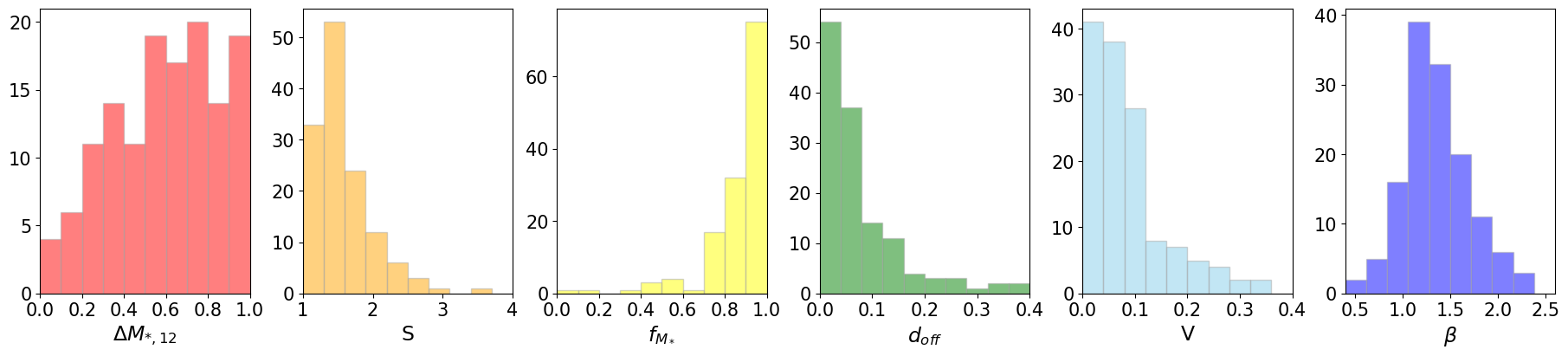}
\caption{Histograms of six dynamical state indicators on (Top) N-cluster Run simulation and (Bottom) HeCS observation data. Top: Black empty, red, green, and blue colored histograms exhibit the number of galaxy clusters for total, recent merger, ancient merger, and relaxed dynamical state, respectively. The overlap of the merger and relaxed sample histogram distribution shows the projection effects on each dynamical state indicator. Bottom: Same indicator histograms as the top panels with different colors. Even though there are slight parameter range differences, the overall parameter distribution seems similar to each other.\label{fig:sim_ind}}
\end{figure*}

Different from our previous study that considered a spectroscopic indicator, we focused solely on photometric dynamical indicators to expand the cluster dataset for this method application. Previously, we utilized four indicators, but we have now introduced two additional indicators based on recent literature. All indicators are affected by the projection effects, which means they are sensitive to the plane of the sky direction. More detailed explanations of the four previously used dynamical indicators can be found in \citet{2024ApJ...970..165K} and the references cited therein. A brief summary of all dynamical indicators is presented in Table \ref{tab:indicator}. Here, we provide brief descriptions of the four established dynamical indicators and detailed explanations of the two new indicators. 

\begin{table}
    \small
	\centering
	\caption{Summary of the dynamical states used in this study. }\label{tab:indicator}
        \begin{tabular}{c|ccc}
         \hline
        \multirow{2}{*}{Indicator}  & Physical & Effective & relaxation \\
          &  parameter & range & criterion  \\
        \hline
                        Sparsity ($S$) & mass & entire & $\simeq$1\\
        Stellar mass gap ($\Delta M_{*, 12}$) & mass & center & $\simeq$0 \\
                   Center offset ($d_{off}$) & position & center & $\simeq$0 \\
        {\scriptsize Satellite stellar mass fraction ($f_{M_*}$)} & mass & entire & $\simeq$0 \\
                Kuiper's V statistics ($V$) & position & entire & $\simeq$0 \\
                Mirror symmetry ($\beta$) & position & entire & Lower\\
         \hline
        \end{tabular} 
        \tablefoot{ The first column describes the physical parameters used for indicators, and the second column shows the region represented by each indicator. The last column displays the relaxation criteria value for indicators.}
\end{table} 

- Sparsity, expressed as
\begin{equation}
S= \Sigma M_*(r_{100})/\Sigma M_*(r_{50})
\end{equation}
where $\Sigma M_*(r_{100})$ and $\Sigma M_*(r_{50})$ are total stellar mass within $100\%$ and $50\%$ of the virial radius of cluster, respectively. The BCG location is the central point in observations, while the center of mass is used in simulations. However, both values are similar since the cluster halo center is treated as the BCG in the simulation. The relaxed state of the cluster is expected to have centrally concentrated galaxy distributions, thus the low sparsity means a relaxed state.

- Stellar mass gap, expressed as 
\begin{equation}
\Delta M_{*, 12}=M_{*, 2nd bcg}/M_{*, bcg}
\end{equation}
where $M_{*, bcg}$ and $M_{*, 2nd bcg}$ are the stellar mass of the BCG and second brightest cluster galaxy, respectively. BCG dominant system means relaxed state, thus the lower value means a relaxed state.

- Center offset, expressed as 
\begin{equation}
d_{off}=|P_{BCG}-P_{M_*\_weighted}|/r_{vir}
\end{equation}
where $P_{BCG}$ is the position of the BCG, and $P_{M_*\_weighted}$ is the stellar mass-weighted center of the cluster member galaxies. For a fair comparison among clusters, we normalize distance with cluster virial radius. Because a merger makes the miscentering of BCG, the small value means the relaxed state. 
\newline

- Satellite stellar mass fraction, expressed as  
\begin{equation}
f_{M_*}=M_{*, sat}/M_{*, cl}
\end{equation}
where $M_{*, sat}$ and $M_{*, cl}$ indicate member galaxy stellar mass and cluster stellar mass, respectively. A massive clump of mass is added to outskirts of main cluster when a merger happens; thus, a small value is expected for the relaxed state. 
\newline

In this study, we further include two different types of spatial asymmetry indicators. Previous observational studies have been considered symmetry of matter distribution as the relaxed state and asymmetry as the disturbed state \citep[e.g.,][]{2010PASJ...62..811O, 2010ApJ...711.1033Z,2020MNRAS.497.5485Y}. We tried to measure these asymmetry features in the galaxy clusters using differences of angular distribution and of mean length from neighbor galaxies.
\newline

- Kuiper's V statistics, used to measure the angular asymmetry of a galaxy distribution based on the difference between the cumulative distribution of a random distribution and a sample distribution, expressed as 

\begin{eqnarray}
V=abs|D^-+D^+|
\end{eqnarray}
where $D^+$ and $D^-$ mean maximum and minimum difference between a random cumulative distribution and a sample cumulative distribution, given by

\begin{eqnarray}
D^-=max[z_{i}-(i-1)/n], \\
D^+=max[i/n-z_{i}]	
\end{eqnarray}
where f is a continuous cumulative distribution function of x, represented by $z_{i}=f(x_i)$.

We measured the angular positions of member galaxies from the BCG by setting north to zero degree and measuring angles clockwise. The cumulative distribution of angles was compared with the random cumulative distribution. The definition and equation are available from \citet{KUIPER196038}.

If member galaxies are located homogeneously (relaxed), the difference between the sample cumulative distribution and the random cumulative distribution will be small. 
\newline

- Mirror symmetry : 
an indicator that also provide information about the positional asymmetry of member galaxy distribution. It measures the mean distance among member galaxies to check the clumpiness of the distribution, given by
\begin{eqnarray}
\beta=<\beta_i>,\\
\beta_i=log_{10}(d’^{(5)}_i/d^{(5)}_i)
\end{eqnarray}
where $d^{(5)}_i$ is the mean distance from galaxy i to the fifth nearest neighbors, and $d’^{(5)}_i$ is the same with $d^{(5)}_i$, but the opposite position of galaxy i through BCG location. If the galaxy distribution is symmetric, the difference between $d^{(5)}_i$ and $d’^{(5)}_i$ is small. Thus, a small $\beta$ value means a symmetric distribution and also means relaxation. However, this indicator is highly sensitive to the membership and completeness of the data (see Section \ref{sec:caveat}). We followed definition and equation from \citet{1988ApJ...327....1W} and \citet{ 2024ApJ...967...14C}.

The histogram distribution of each indicator from simulation data is shown in the top panels of Figure \ref{fig:sim_ind}, while the same indicators from observation data are shown in the bottom panels of Figure \ref{fig:sim_ind}. Merging and relaxed clusters (plotted in different colors) are not well separated in any of the indicators due to projection effects. Although the histogram distributions are not perfectly matched between simulation and observation, this is caused from size difference between two data sets (see Figure \ref{fig:sim_ind}).

\subsection{Bayesian classifier and class conditional distributions with a GMM}\label{sec:NSgmm}
We utilized a Bayesian classifier where class conditional distributions are estimated through an infinite GMM, to address the limitations of our previous study. In our earlier methods, we were unable to consider parameter spaces larger than four dimensions and could not provide specific probability information for more detailed merger stages. However, with the approach proposed in this section as a classifier, we can model each merger stage in the 6D space of indicators. Additionally, it is possible to project the 6D class-conditional distributions into lower dimensional indicator spaces.

\subsubsection{Bayesian classifier}\label{sec:bayes}
We divided the simulation sample into training and test datasets, using a 70:30 ratio, stratified by class (merger stage). The training data were employed to train and build class-conditional distributions, while the test data were used to assess the classifier's performance. 

The Bayes classifier is based on the estimation of class-conditional distributions $p(\mathbf{x}|C_k)$, where the distribution of points belong to class $C_k$. While generally considered the optimal classifier, its performance relies heavily on the quality of the class-conditional distributions in practice. 

The naive implementation of the Bayes classifier assumes conditional independence across features; however, this assumption disregards correlations in the dataset that might be meaningful in informing the classification. However, much more nuanced estimations of the class-conditional distributions might be obtained, given a sufficiently rich dataset. Thus, we did not use the naive implementation in this work.

Given $K$ classes $C_k$, it is possible to estimate the posterior distribution, the probability that the x is originated from class  $C_k$, $p(C_k|\mathbf{x})$ using the Bayesian rule of
\begin{equation}\label{eq:Bayes}
p(C_k|\mathbf{x}) = \frac{p(\mathbf{x}|C_k) p(C_k)}{p(\mathbf{x})} = \frac{p(\mathbf{x}|C_k) p(C_k)}{\sum_{i=1}^K p(\mathbf{x}|C_i) p(C_i)} ,
\end{equation}
where $p(\mathbf{x}|C_k)$ is the likelihood of the data given the model for class $C_k$, $p(C_k)$ the prior over class $C_k$ and the denominator acts as a normalization (often referred to as the model evidence).

Given a new sample $\mathbf{x}^*$ then, the above equation assigns a probability for that point to have originated from class $C_k$. If a hard assignment needs to be performed, the parent class (estimated label) is estimated as
\begin{equation}\label{eq:classifier}
\hat C = \arg_{k} \max p(C_k|\mathbf{x}^*).
\end{equation}

Once the Bayesian classifier is adopted, we only need to specify a framework for the estimation of the class-conditional distributions. This is reported in the following section.

\subsubsection{Class-conditional distributions}\label{sec:classc}
At the core of the proposed methodologies usually lies the Gaussian Mixture Model (GMM). 

As it is particularly useful because it leverages the flexibility and regularities of the Gaussian distribution. The probability density function (PDF) of a GMM can be written as
\begin{equation}
p(\mathbf{x}|\boldsymbol{\Theta}) = \sum_{k=1}^M \pi_k p(\mathbf{x}|\boldsymbol{\theta}_k),
\end{equation}\label{eq:pdf}
where $p(\mathbf{x}|\boldsymbol{\theta}_k) = \mathcal{N}(\boldsymbol{\theta}_j)$, probability of x can be described with a normal distribution defined by parameter vector $\boldsymbol{\theta}_k = [\boldsymbol{\mu}_k, \Sigma_k]$(i.e., the mean and covariance matrix of mixture component, $k$, respectively). Here, $\boldsymbol{\Theta} = [\boldsymbol{\theta}_1,\dots,\boldsymbol{\theta}_M]$ contains parameters of all components. The parameter $\pi_k$ is the mixture coefficient for a component, $k$, of the mixture and must satisfy
\begin{equation}
\pi_k \geq 0, ~ \forall k=1,\dots,M; \qquad \sum_{k=1}^M \pi_k = 1. 
\end{equation}
When the mixture coefficient for component \textit{k} is larger than zero and for all \textit{k} is part of \textit{M}, the summation of mixture coefficients should be one. 
In both equations, the number of components in the mixture is $M$.

When estimating the density of a dataset with GMM, the main hyperparameter to be identified is the number of components, $M$. Different routines exist to estimate it directly from the data, leveraging information theoretic quantities such as the Bayesian or Akaike information criteria (BIC or AIC). However, estimation of optimal hyperparameter and parameters is a data-intensive process that requires multiple splits of the set, resulting in possible biases.

An alternative to classical GMM is the Infinite GMM \citep{Rasmussen_1999}. This is a Bayesian formulation of the GMM that imposes priors on the parameters, $\boldsymbol{\Theta}$, of the mixture components and the corresponding hyperparameters. The net effect of this formulation is that the estimation of the effective number of components in the mixture is delegated to an approximation of a Dirichlet Process. 

In practice, only the maximum number of components and the concentration prior of the Dirichlet process need to be specified to constitute hyperparameters for the Dirichlet process-GMM (DP-GMM). Higher values of the concentration prior will enforce a higher number of components with large mixture coefficients and vice versa. It is worth noting that the quality of the obtained density is not necessarily affected by the number of components, but its complexity is. We adopted the version implemented in the \textit{python sklearn} library \citep{2011JMLR...12.2825P}.

In summary, the class-conditional distribution for each merger stage, $p(\boldsymbol{x}|\boldsymbol{\theta}_k)$ has been constructed using a DP-GMM, testing various parameters and identifying the optimal one via out-of-sample likelihood. Using equation \ref{eq:Bayes}, we obtained the corresponding posterior $p(C_k|\boldsymbol{x})$, and we calculated the accuracy by assigning unseen data (test set) to the most likely merger stage (class) via eq. \ref{eq:classifier}.

The best model classifier was created for each merger stage, developed for two and three merger stage classifications, and was subsequently applied to the observational data under the same merger stage conditions.

\subsubsection{Confusion matrix and classification report}
Because we divided the sample into training data and test data, we can judge the performance of the classifier with precision, accuracy, and recall values. We present a brief set of definitions with equations for each quantifying values below. Here, capital letters T, F, N, and P represent true, false, negative, and positive, respectively. 
True and false represent the actual class of test data, while positive and negative represent the predicted class of test data. These concepts are used to measure the performance of machine-learning classification by combining concepts of actual and predicted classes, such as true-positive (TP), false-positive (FP), and false-negative (FN). For example, when the classifier predicts a cluster as relaxed, when it is actually in a relaxed state, it counts as a TP classification result. However, if it is not an actual relaxed state, it counts as FP and so on. 
\newline

-Precision : to show how well the model predicts the positive class, we use 
\begin{eqnarray}
Precision=\frac{TP}{TP+FP}
\end{eqnarray}
-Accuracy : to measure how well the model predicts the results, we use
\begin{eqnarray}
Accuracy= \frac{TP+TN}{TP+TF+FP+FN}
\end{eqnarray}
-Recall : to measure how often the model correctly identifies true positives from all the actual positive samples, we use
\begin{eqnarray}
Recall=\frac{TP}{TP+FN}
\end{eqnarray}

Precision can represent completeness of classification, while accuracy indicates the purity of the classified results. Here, we used average per-class accuracy to quantify performance,

\begin{eqnarray}
\mathsmaller{\text{Average\:per-class\:accuracy} =
\frac{\text{binary\:accuracy\:for\:each\:class\:label}}{\text{number\:of\:class}}}
\end{eqnarray}
By comparing these parameters, we could find the best model classifiers for each merger stage. We selected the training and test data using random bootstrapping resampling to minimize bias from the sample distribution. Thus, the precision, recall, and accuracy values for each model were compared as the mean of the iterated random resampling. 

\section{Analysis and result}\label{sec:result}

In this section, we present the results of our new classification method for both two-merger and three dynamical state classifications. First, we describe the best model condition, show the results from projected classifier applications by different numbers of indicator, and give the best indicator combination results. We also compare the performance of precision, accuracy, and recall results with those from a non-projection model. Finally, we demonstrate the application of this method to observational data and present the classification results.

\subsection{Best model analysis}\label{subsec:bestmodel}
\begin{figure*}
\centering
\includegraphics[scale=.27]{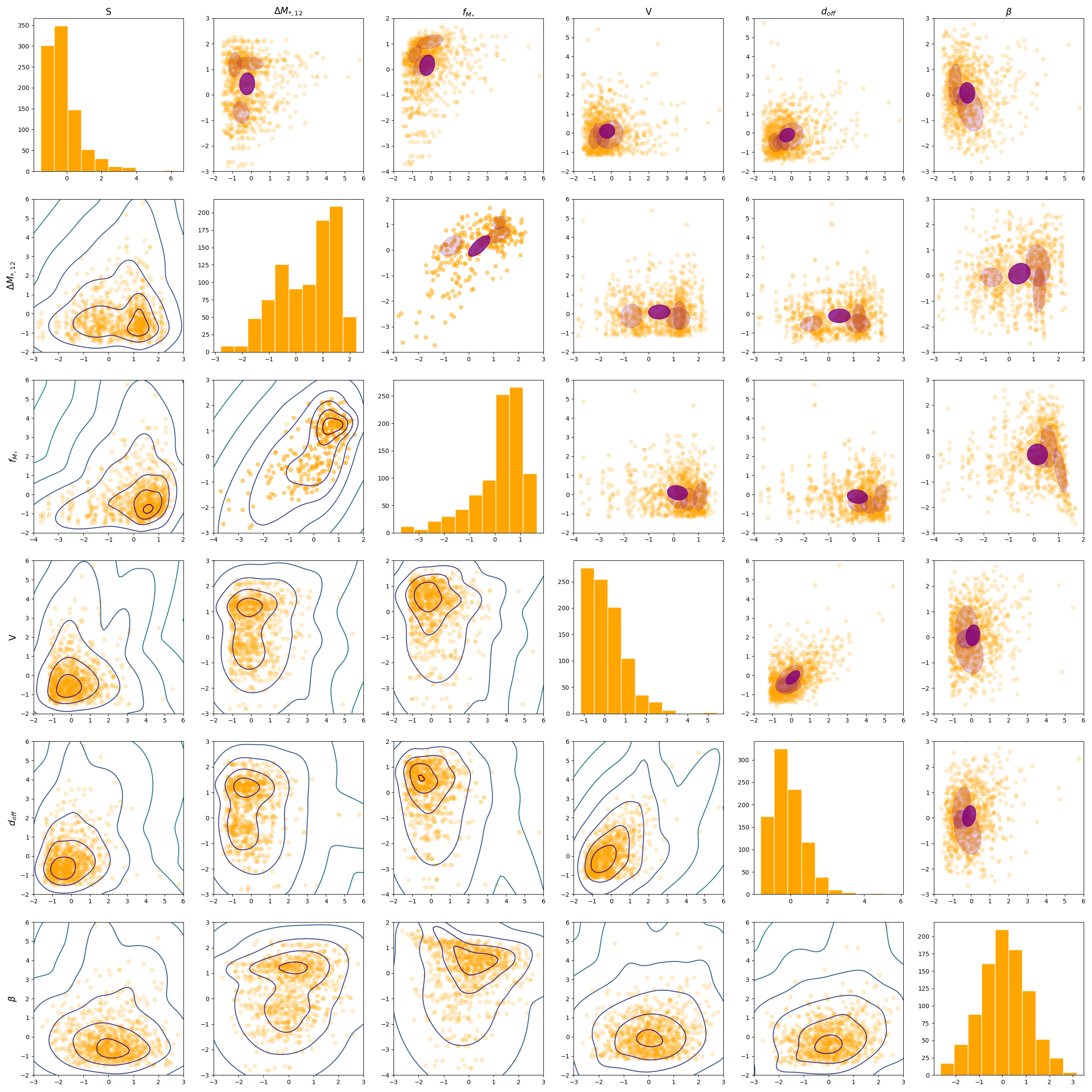} 
\caption{2D Correlation plots for each indicator distribution by merger sample. The bottom-left corner and top-right corners give the scatter plots of the indicator against the other. The diagonal shows the histogram of individual indicators. In the bottom-left panels, we overplot probability contours (varying probability value in each panel). In the top-right panels, we overplot ellipses indicating the shape of the Gaussian for all components with a weight concentration prior > 0.1. We note that ellipses match the location of the high probability contours.} \label{fig:modeling} 
\end{figure*}

Because the distribution of the merger stage samples of each indicator resembles a shell structure, which makes it difficult to separate subpopulations, we selected the GMM for this highly overlapped class distribution. We made Gaussian models for each merger stage sample, as illustrated in Figure \ref{fig:modeling}. We then combined the probabilities from each GMM to obtain the class-conditional probability distribution.

\begin{figure*}
\centering
\includegraphics[scale=.27]{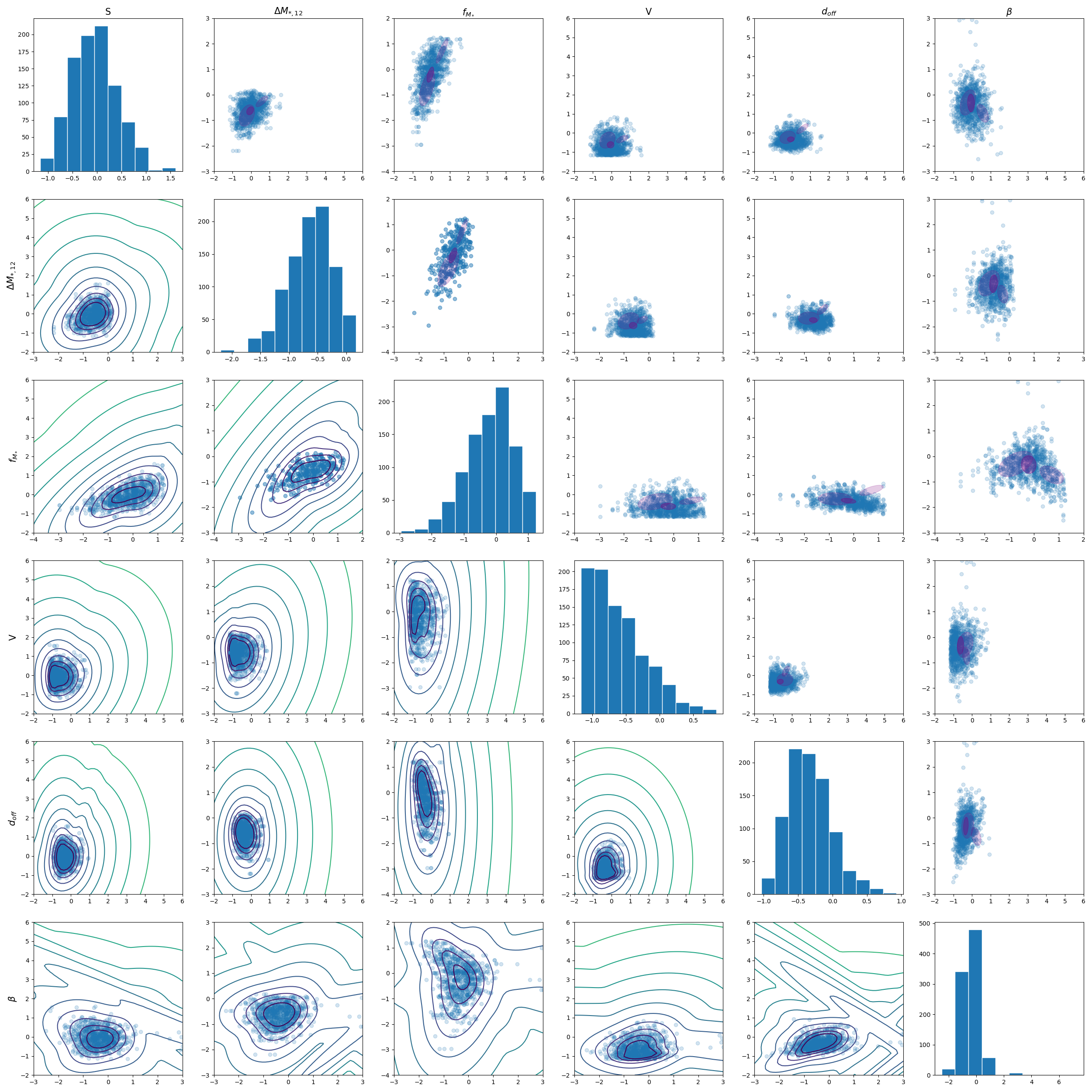} 
\caption{2D Correlation plots for each indicator distribution by relaxed state sample. Instructions are the same as Figure\ref{fig:modeling}. }\label{fig:modeling1} 
\end{figure*}

By testing the full covariance matrix, we analyzed the 6D correlations simultaneously. To improve the overall understanding, we visualized this data using 2D correlation plots. Figure \ref{fig:modeling} and Figure \ref{fig:modeling1} illustrate the 2D distributions for each dynamical indicator across different merger stages. The upper 2D parameter spaces show the position of highly weighted Gaussian components of parameter distributions with purple ellipses, used to model the parameter distribution. The histograms in diagonal panels provide the distribution of each indicator, while the lower 2D parameter spaces show the probability contours derived from the GMM. 

To properly and efficiently model the distribution using the Gaussian formula, we need several hyperparameters, including random values for the starting point (e.g., $\mu_k$, $\Sigma_k$), a tolerance value to exit the loop, and a weight concentration prior parameter to assign weights to the Gaussian distributions. We chose the Dirichlet process for the weight concentration prior type because (while the Dirichlet distribution models probability with a fixed number of components) the Dirichlet process determines the optimal number of components itself. The Dirichlet process modifies the coefficients and gradually reduces the ones that reflect less variance in the data. At the end of the training, due to the Dirichlet process, the number of mixture coefficients that are effectively meaningful for the construction of the model is generally much lower than the total number of components in the mixture.
However, when we check the histogram of weights, more than 80$\%$ of the weights are valued at less than 0.01. It shows that although the Dirichlet process reduces some of the lowest weight components, our model needs those lower weight components to describe the entire distribution.

We tested all the hyperparameters, along with the number of Gaussians made the most meaningful probability differences. Thus, we mainly tested various combinations of the number of Gaussian fittings to find the best model. Here, we mention that the hyperparameter labeled as the "number of components" is the maximum number of Gaussian components used in the Dirichlet process. Also, Bayesian priors with equal percentages or sample-proportional percentages lead to negligible differences. 

\begin{figure*}
\centering
\includegraphics[scale=.44]{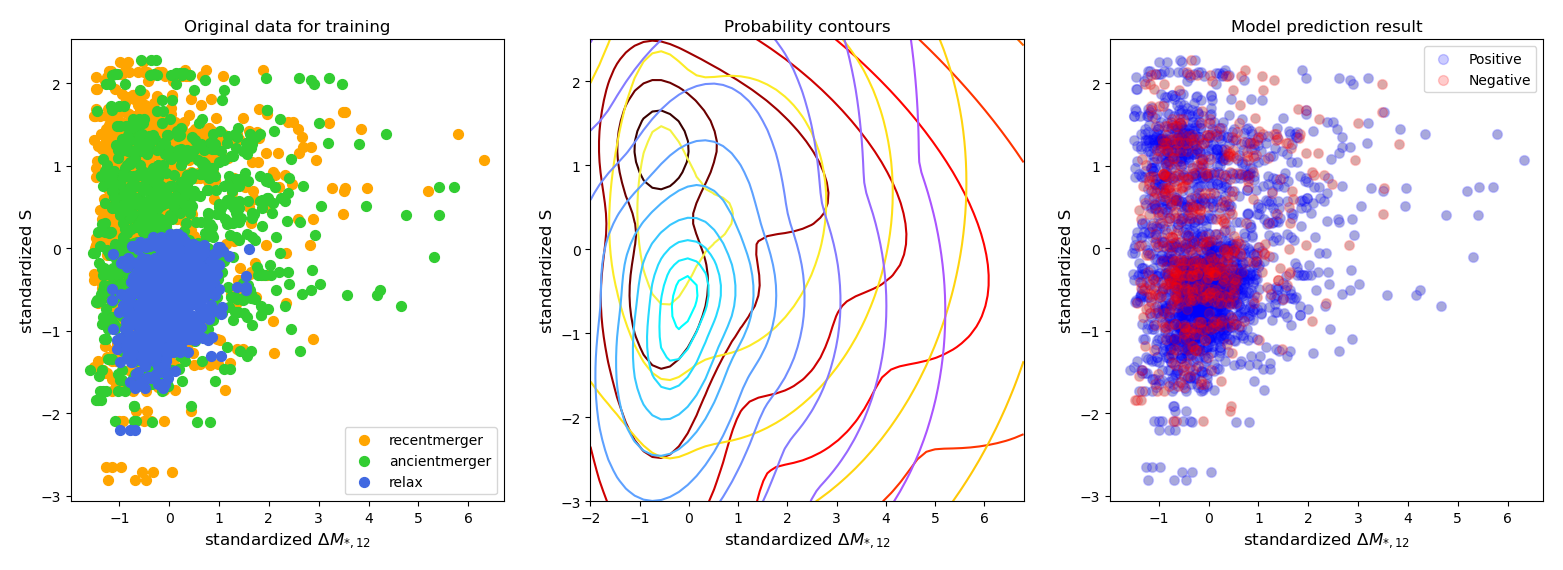} 
\caption{Example figures of the classification process and result from the 6-indicator Bayesian classifier. Figures are 2D, but the actual classifier was made in a 6D indicator space. Left: Standardized original simulation data distribution on 2D indicator space. Merger stages are shown with different colors, and check it in the legend. Middle: Probability contours. Red, yellow, and blue lines represent probability contours of recent merger, ancient merger, and relaxed samples. Based on these probability contours, the model classifies the dynamical state as shown in the middle panel. Right: Prediction result of the overall sample distribution. Blue dots show the matched classification, and red dots show the mismatched classification. The center of the blue contour region shows good classification, but the center of the red contour region shows conflict of classifications between the recent merger and the ancient merger classification results. It represents an uncertain region of classification. }\label{fig:expsim} 
\end{figure*}

The probability contours for each merger stage are displayed in the middle panel of Figure \ref{fig:expsim}. Each probability contour was created based on the distribution of each merger stage, as shown in the left panel of the same figure. The classification results are presented in the right panel.

To mitigate specific sample distribution bias, probability contours for each subsample were created with the mean distribution of subsamples from the bootstrap resampling method, even though we used the full sample distribution for the Figure \ref{fig:expsim}. We resampled each merger stage by maintaining the same percentage as the entire merger stage sample ratio. 

To compare the performance across models using different numbers of Gaussians, we analyzed the mean values of precision, recall, and accuracy of the models for each subsample. The differences among the resampled subsamples were less than 0.01 in probability values, which is smaller than the 0.1 probability difference identified when varying the number of components.

We evaluated the classifiers based on the average values of precision, recall, and accuracy. The results indicate that the two-merger stage classifier achieves an average performance of 92$\%$ in accurately reproducing the true class, while the three-merger stage classifier demonstrates an average performance of 77$\%$ in reproducing the true class. The best model conditions for the two and three dynamical state are presented in Table \ref{tab:bestmodel}. 

\begin{table}
\small
\centering
\caption{Six indicator best model precision, recall, and accuracy values by number of merger stages.}\label{tab:bestmodel}
\begin{tabular}{c|cc|ccc}
\hline
\multirow{2}{*}{merger stage} & \multicolumn{2}{c|}{2}    & \multicolumn{3}{c}{3}   \\ \cline{2-6} 
                  & \multicolumn{1}{c|}{merger} & relaxed& \multicolumn{1}{c|}{recent} & \multicolumn{1}{c|}{ancient} & relaxed\\ \hline
                Numb of Gauss  & \multicolumn{1}{c|}{150} & 150 & \multicolumn{1}{c|}{100} & \multicolumn{1}{c|}{100} & 50 \\ \hline
               Precision & \multicolumn{1}{c|}{0.9516} &  0.8938& \multicolumn{1}{c|}{0.7123} & \multicolumn{1}{c|}{0.7176} & 0.9157 \\ \hline
                Recall  & \multicolumn{1}{c|}{0.9462} & 0.9025 & \multicolumn{1}{c|}{0.7601} & \multicolumn{1}{c|}{0.6989} & 0.8785 \\ \hline
                 Accuracy & \multicolumn{2}{c|}{0.9244}    & \multicolumn{3}{c}{0.7816} \\ \hline
                 Average & \multicolumn{2}{c|}{0.9252}    & \multicolumn{3}{c}{0.7798} \\ \hline
\end{tabular}
\tablefoot{ Numb of Gauss shows the number of Gaussian components used for model indicator distributions.}
\end{table}


\begin{SCfigure*}
\centering
\includegraphics[scale=.44]{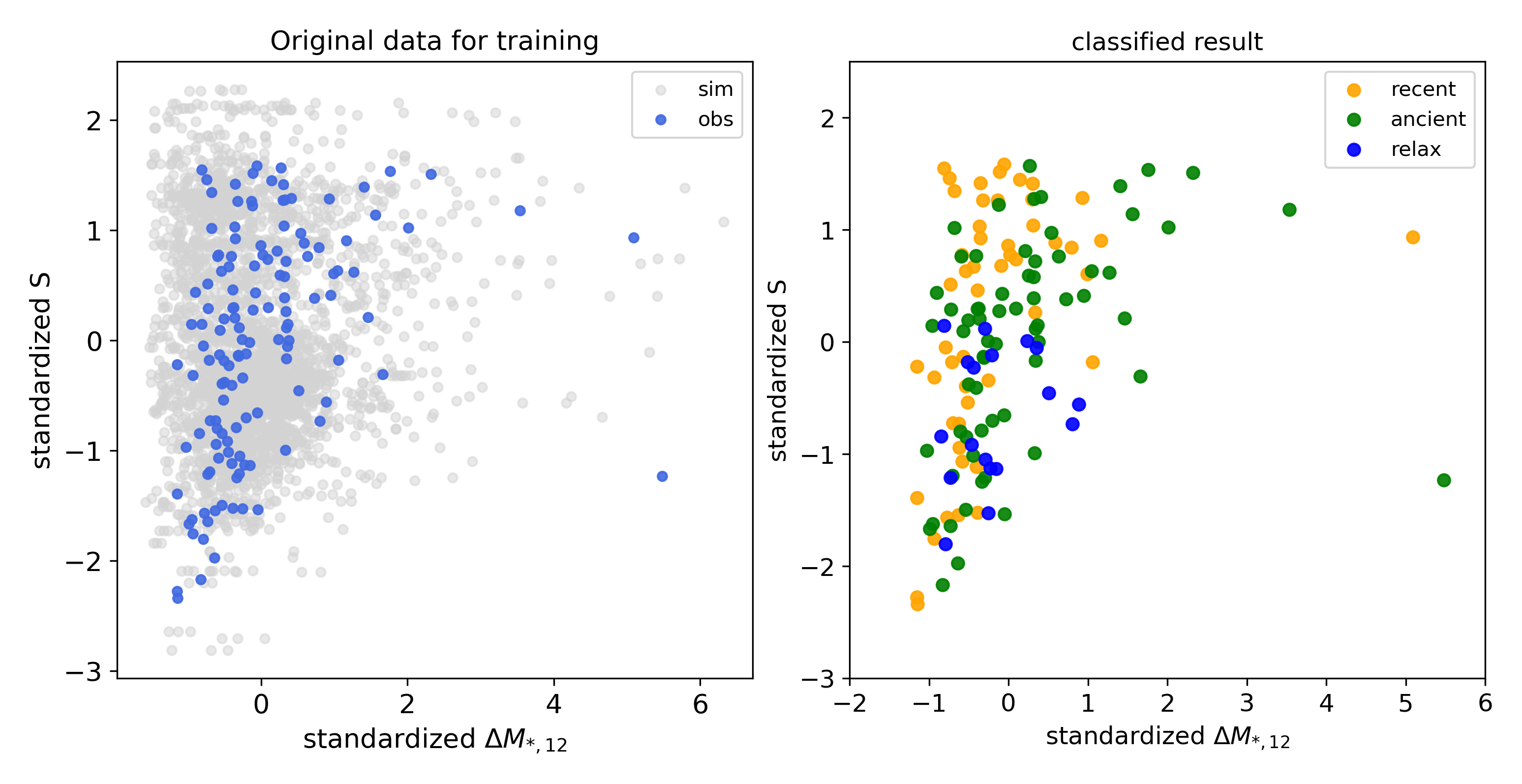} 
\caption{Example figures for a modeled Bayesian classifier application on the observation data. Again, figures are two dimensional, but the actual classifier was made in a 6D indicator space. Left: Original 2D observation data distribution. Classifier was modeled using simulation data distribution (white), and then the trained classifier is applied to observation data (Blue). Right: classified results of observation data. Each color represents different merger stages, and check it in legend.}\label{fig:expobs} 
\end{SCfigure*}
\subsection{Application on observation data}
\begin{table}
\centering
\small
\caption{HeCS cluster classification results from two and three merger stage classifiers.} \label{tab:number}
\begin{tabular}{c|cc|ccc}
\hline
\multirow{2}{*}{merger stage} & \multicolumn{2}{c|}{2}    & \multicolumn{3}{c}{3}  \\ \cline{2-6} 
                  & \multicolumn{1}{c|}{merger} & relaxed & \multicolumn{1}{c|}{recent} & \multicolumn{1}{c|}{ancient} & relaxed \\ \hline
              Numb of cl    & \multicolumn{1}{c|}{116 (105)} & 18 (8)& \multicolumn{1}{c|}{58 (18)} & \multicolumn{1}{c|}{52 (12)} & 25 (4) \\ \hline
               $\%$   & \multicolumn{1}{c|}{81} & 19 & \multicolumn{1}{c|}{48} & \multicolumn{1}{c|}{46} & 6 \\ \hline
\end{tabular}
\tablefoot{ Numb of cl shows the number of clusters. Numbers in the parentheses represent the number of clusters having a higher than 90$\%$ probability.}
\end{table}

We applied this cross-conditional probability distribution to the HeCS cluster catalog data. As shown in the left panel of Figure \ref{fig:expobs}, our classifier was modeled with simulation train data (white), then applied to observation data (blue). We applied probability contours (as shown in the middle panel of Figure \ref{fig:modeling}) to the observation data and could get the classification result (as in the right panel). Once the classifier has been modeled, applying it to observational data is quick and straightforward.

\begin{figure*}
\centering
\includegraphics[scale=.7]{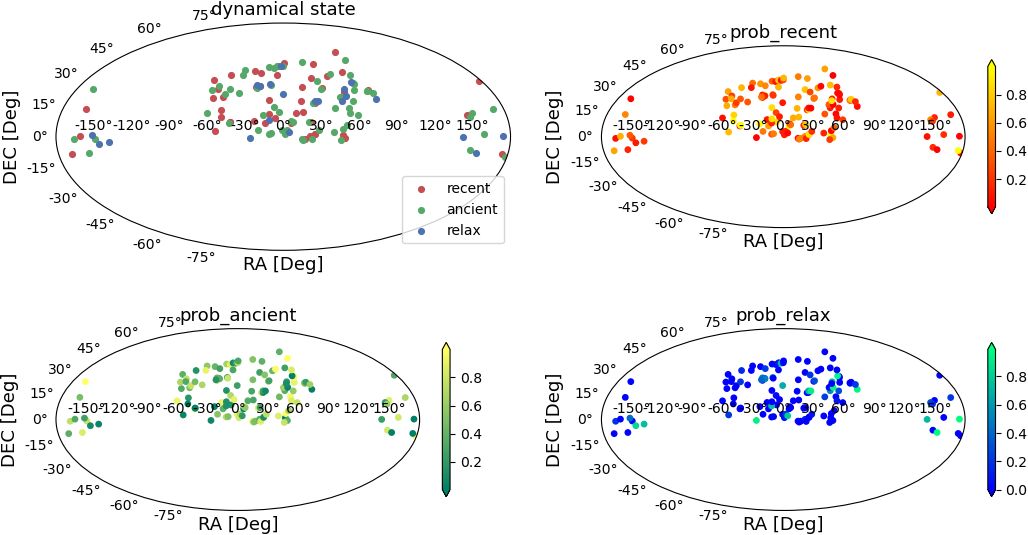} 
\caption{Spatial distributions of 135 classified observation clusters. Top-left: Dichotomy classification result. Red, green, and blue circles represent recent merger stage, ancient merger stage, and relaxed state clusters. Spatial distributions of observation clusters. From the top-right to the bottom of figure, the color gradient shows probabilities of recent merger (red, top-right), ancient merger (green, bottom-left), and relaxed state (blue, bottom-right). }\label{fig:dyst} 
\end{figure*}

Figure \ref{fig:dyst} represents the spatial distribution of the dynamical state of HeCS clusters as determined by the three-merger stage classifier. Each class is assigned based on which class has the highest probability relative to the others (see Appendix \ref{app:2}). However, each cluster has its own set of probabilities for all merger stages, allowing us to examine these probabilities as percentages. 

To display the tendencies, we assigned a dynamical state with the highest probability to each cluster. However, there are many alternative ways to use these probability values to assign a dynamical state. For example, we could define a high confidence sample, where only objects with greater than 90$\%$ probability are included. Alternatively, instead of assigning individual objects to specific categories, we can use the probabilities as weights. For instance, measuring specific properties of all the clusters (e.g., the blue fraction), but also weighting their contribution to the measurement by the merger or relaxed probability. This approach fully avoids making an arbitrary choice for how to classify the clusters and it is an especially useful approach when the number statistics of clusters is low and splitting them into separate categories could reduce the number statistic further. Therefore, the fact that the model provides probabilities is flexible and should be considered a strength of the approach. Figure \ref{fig:dyst} displays the probabilities of each merger stage using different colors. The color bars indicate the magnitude of these probabilities.

Based on the results in Table 4, the two-merger stage classifier identifies a similar number of merger classifications as the three-merger stage classifier, which means our classifier works in a consistent way. Having a large fraction of merging clusters in the observed sample could arise because of the way the sample is selected. The HeCS cluster sample was selected from X-ray dominant and massive clusters. And merging clusters have a tendency towards larger masses than relaxed clusters, following the known mass-dynamical state dependency \citep{2012MNRAS.427.1322L, 2019ApJ...887..264R, 2021A&A...652A.155S}. 

Although the absolute number and percentage of merger samples are similar for the two classifications, with 116 samples (85$\%$) from the two-merger stage classifier and 111 samples (82$\%$) from the three-merger stage classifier, there is a significant difference in the number of high-probability samples. In particular, there are 105 for the two-merger stage classifier compared to only 30 for the three-merger stage classifier. 

This discrepancy can be further understood by examining the precision, recall, and accuracy values shown in Table \ref{tab:bestmodel}. The more specific classification makes detection more challenging due to a reduced training dataset, which results in the three-merger stage classifier having lower precision, recall, and accuracy values, as well as a smaller number of high-probability samples.

\subsection{Model projection and number of indicator effects} \label{sec:proj} 

The Bayesian classifier has the advantage of having class-conditional distributions built in the high dimensional space. In the inference phase, the class-conditional distributions can be projected on the lower dimensional space spanned by the available indicators. The projections are more informative than the GMM built on the corresponding low dimensional space, carrying the high dimensional correlations through the projections. For instance, even with just two indicator variables, we can classify dynamical states using information from six indicator spaces. Therefore, we projected classifiers for two-merger and three-merger stages into the lower number of indicator spaces to evaluate their performance (e.g., train six-indicator information to classify 5, 4, 3, and 2 indicator classification). We compared these results with those from non-projected classifiers, which utilized the same amount of indicator information to train (e.g., trained two-indicator information to classify a two-indicator classification).

\begin{figure*}
\centering
\includegraphics[scale=.52]{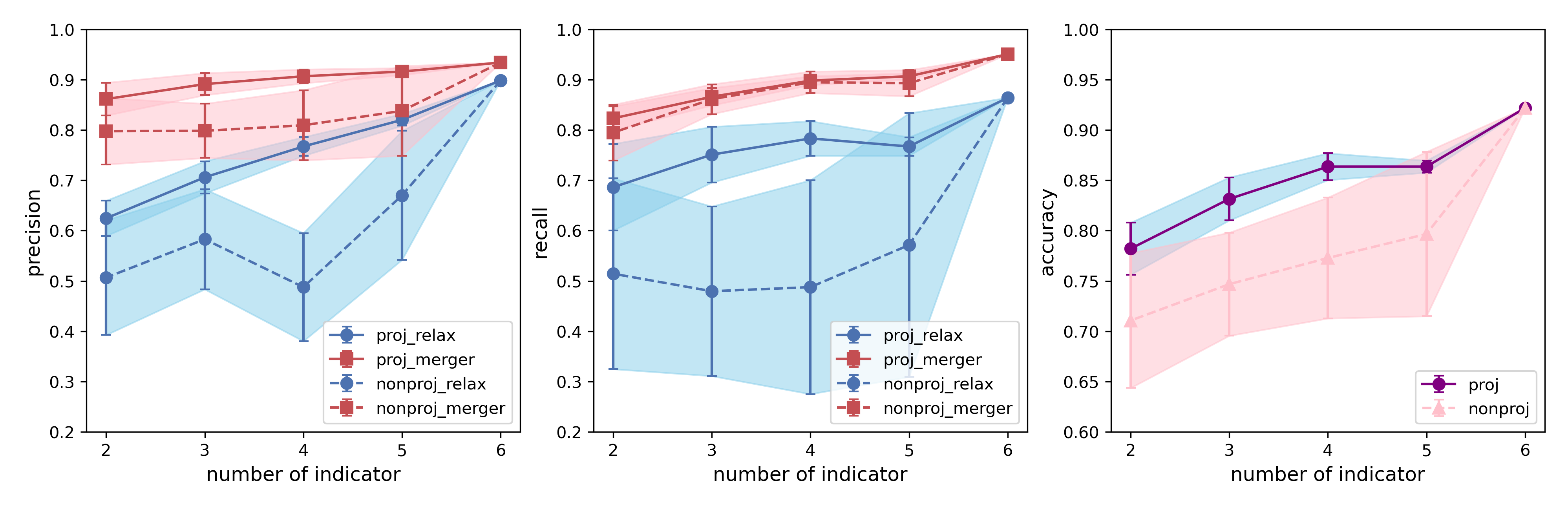} \\
\includegraphics[scale=.46]{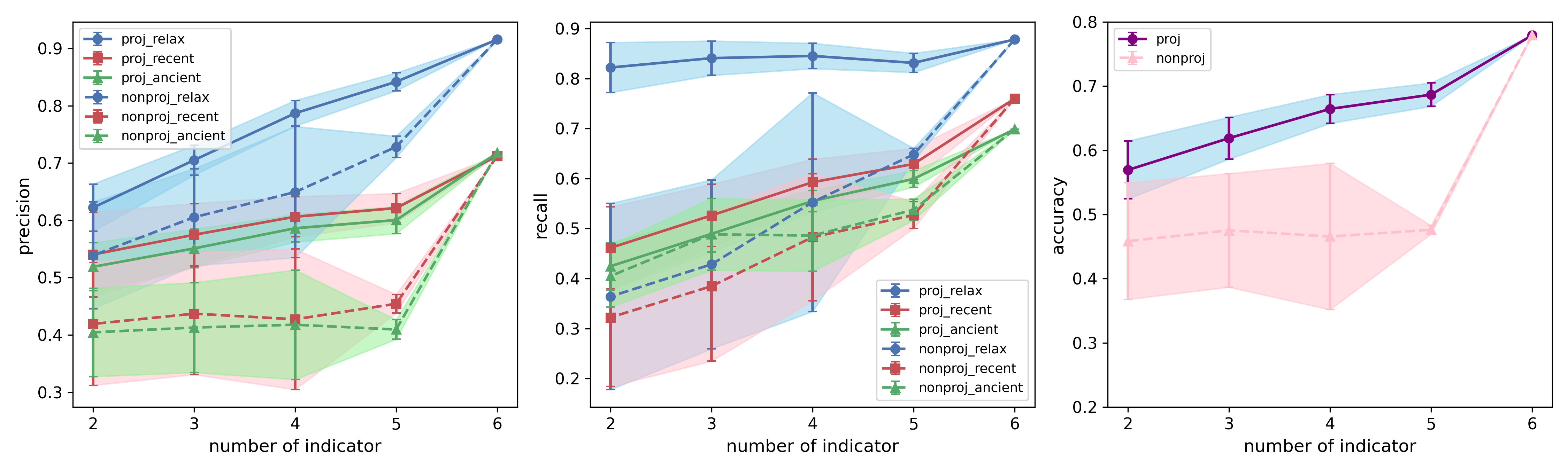} 
\caption{Precision, recall, and accuracy changes by number of combined indicators. The upper three plots show results from two merger stage classifications (merger and relax), and the lower three plots show results from three merger stage classifications. From Left to right columns, precision, recall, and accuracy results were shown. Each color of line represents a different merger stage, which can be identified in the legend. The solid line and dashed line indicate the projected classifier and non-projected classifier results, respectively. Shade shows scatter from different indicator combinations. }\label{fig:PRA} 
\end{figure*}

Figure \ref{fig:PRA} presents the results for precision, recall, and accuracy based on different numbers of indicator information used. The shaded area illustrates the variation of values from various indicator combinations, highlighting the largest error value compared to other errors. Notably, the projected classifier results (represented by the solid line) consistently outperform the non-projected classifier results (depicted by the dashed line) in both two- and three-merger stage classifications, which means the projected classifier can provide better classification results.

In terms of precision, the projected classifier shows an average improvement of 10$\%$ over non-projected classifier results. The recall parameter has the largest 40 $\%$ increase percentage for the relaxed dynamical state sample. Additionally, the accuracy of the projected classifier demonstrates about a 10$\%$ better performance, compared to the non-projected results.

In addition, we can revisit the finding that a larger number of indicator combinations leads to better results. As a byproduct of projection analysis, we can get the best combination results for different numbers of indicator combinations. The best combination, based on the number of indicators, is ranked by magnitude difference, center offset, sparsity, Kuiper V, mirror asymmetry, and satellite stellar mass fraction as shown in Table \ref{tab:bestcombi}. The order of indicators was chosen based on their overall performance in terms of precision, recall, and accuracy in classification. We discuss the importance rank of the indicator in Section \ref{subsec:bestind}.

\begin{SCfigure*}
\centering
    \begin{subfigure}[b]{0.33\textwidth}
        \includegraphics[width=\textwidth]{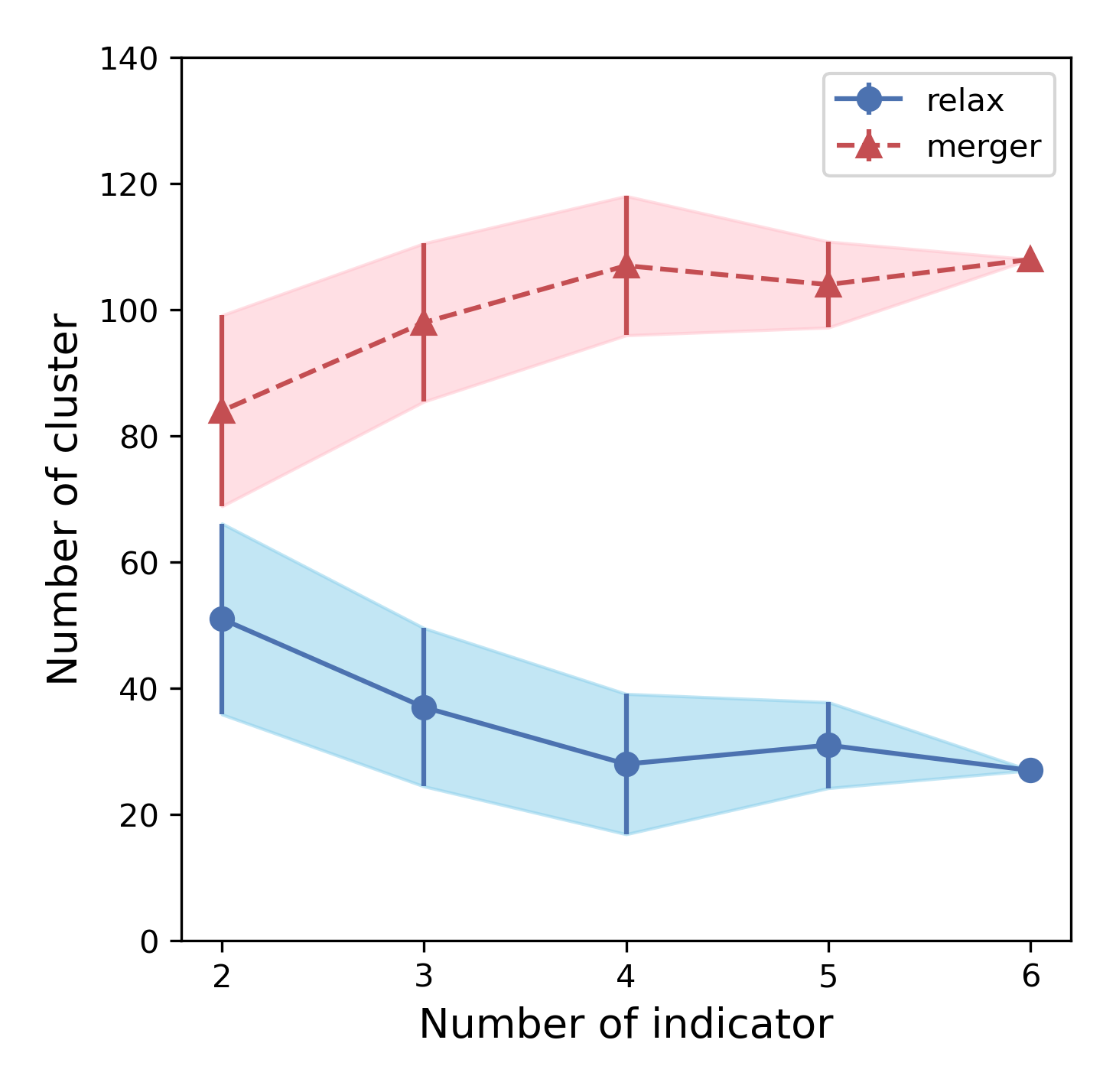}
        
    \end{subfigure}
    \hfill
    \begin{subfigure}[b]{0.33\textwidth}
        \includegraphics[width=\textwidth]{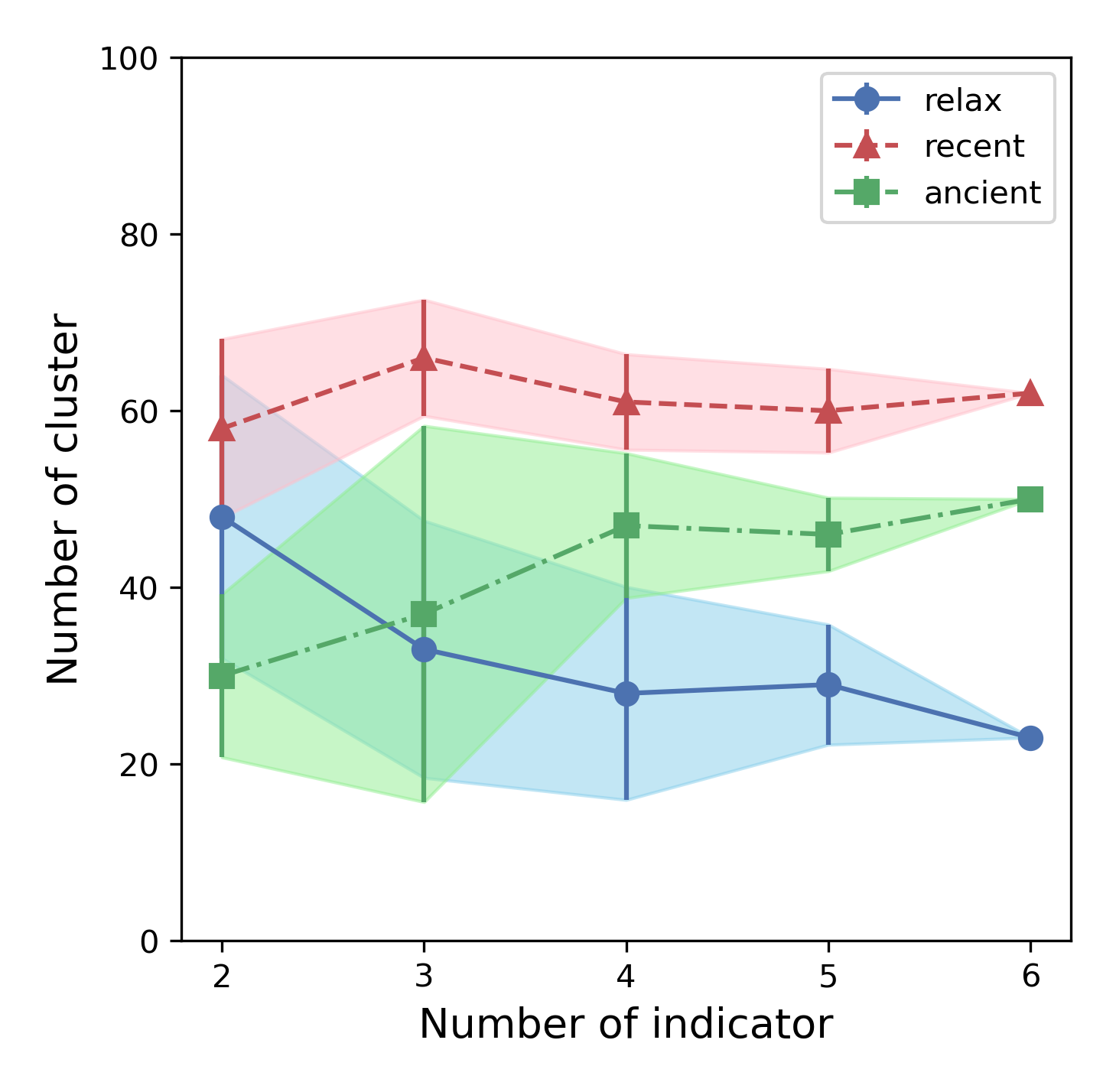}
        
    \end{subfigure}
\caption{Classification number of cluster tendency by number of combined indicators. Left: Projected model results for two merger stage classifications and the Right: Projection model results for three merger stage classifications. Different colors, symbols, and line styles represent different merger stages, and it can be checked in the legend. }\label{fig:number} 
\end{SCfigure*}

Additionally, we applied projected classifiers to the HeCS cluster data. Figure \ref{fig:number} illustrates the classification tendencies based on the number of classified clusters. By an increase in the number of indicators, both the two-merger and three-dynamical state classifiers show a decrease in the number of relaxed clusters while indicating an increase in the number of merger clusters. These results appear consistent with the six-indicator classifier result.

The fraction of recent mergers remains roughly constant, while the curves that vary (and even end up inverted) characterize the relaxed and ancient mergers. This indicates that the ancient mergers represent an intermediate state, which can be mistaken for the relaxed cluster when using only a few components.

\begin{table}
	\centering
	\caption{Best indicator combination by number of combined indicators. }	\label{tab:bestcombi}
	\begin{tabular}{c|c} % four columns, alignment for each
		\hline
            Number of ind & Best combination\\
            \hline
           2& $\Delta M_{*, 12}$, $d_{off}$\\
           3& $\Delta M_{*, 12}$, $d_{off}$, S\\
           4& $\Delta M_{*, 12}$, $d_{off}$, S, V\\
           5& $\Delta M_{*, 12}$, $d_{off}$, S, V, $\beta$\\
		\hline
	\end{tabular}\label{tab:xtrain}
    \tablefoot{ Indicators are represented with the acronym. Information about the indicator and acronym is in the section \ref{sec:ind}.}
\end{table}

\section{Discussion}\label{sec:discussion}
In this section, we discussed indicator dependency by mass and redshift and correlation among the indicators. Furthermore, we present the comparison result with our previous method and this study. Some caveats of Bayesian classification are also discussed below.
\subsection{Indicator dependency on mass and redshift}\label{sec:dependency}
\begin{SCfigure*}
\centering
    \begin{subfigure}[b]{0.38\textwidth}
        \includegraphics[width=\textwidth]{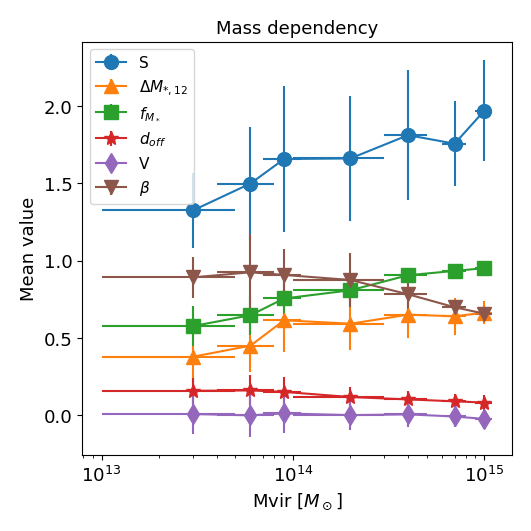}
        
    \end{subfigure}
    \hfill
    \begin{subfigure}[b]{0.38\textwidth}
        \includegraphics[width=\textwidth]{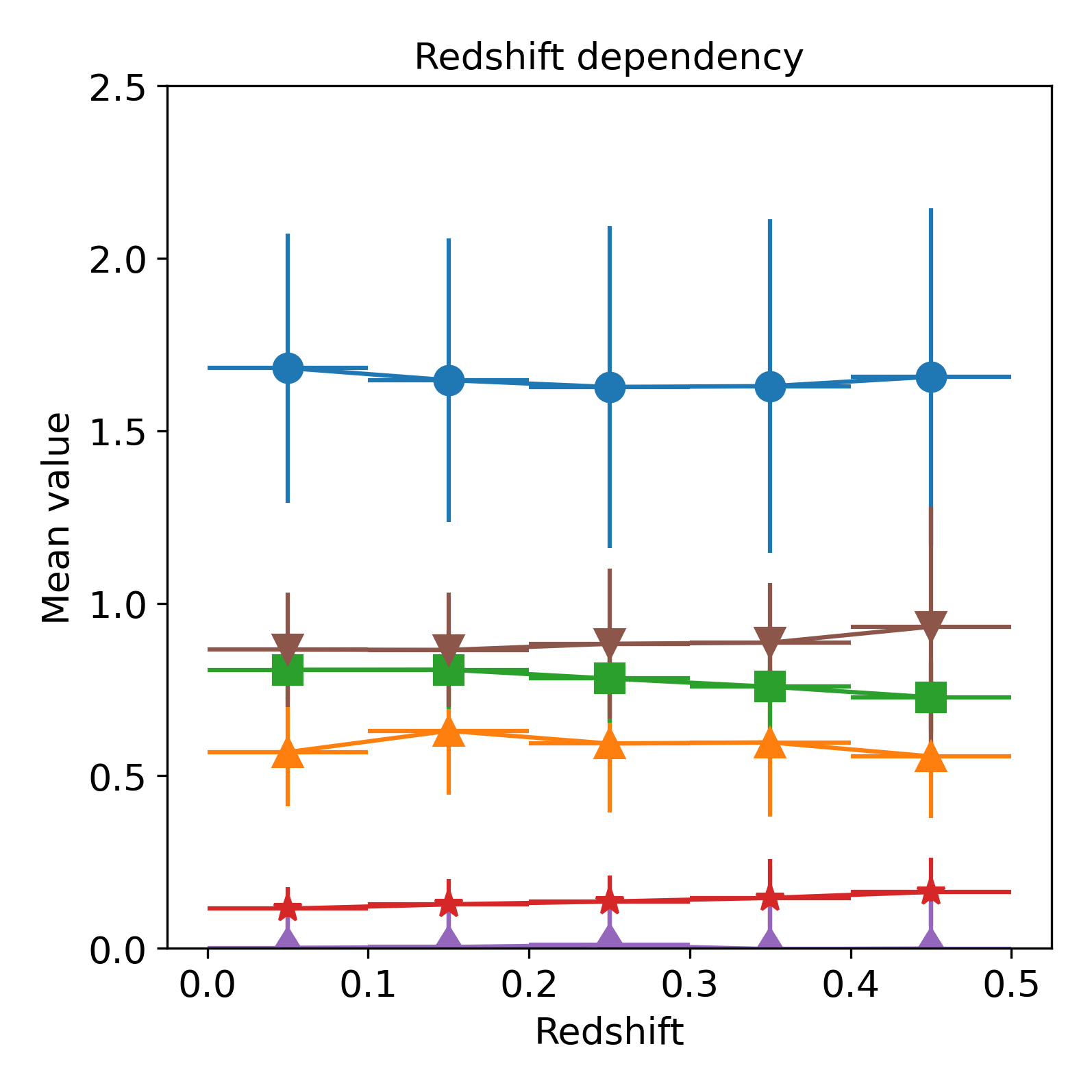}
        
    \end{subfigure}
\caption{Left: Mass dependency of six dynamical state indicators. Each color and shape of symbols represent different dynamical state indicators. Sort of indicator can check in the legend. Detailed information of indicators is in section \ref{sec:ind}. X error shows bin size and y error shows standard deviation of values within bin. Right: Redshift dependency of six dynamical state indicators. Features are same with upper panel.}\label{fig:massdepend} 
\end{SCfigure*}
 
We tested dependencies of dynamical indicators from simulation data by mass and redshift. For the mass dependency test, we divided the cluster mass range $2\times 10^{13}\sim2\times 10^{15} M_\odot$ into seven mass bins (see left panel in the Figure \ref{fig:massdepend}). All indicators show distinct increasing or decreasing trends and the indicator exhibiting the widest variance was sparsity. It is increased from 1.33 to 1.97, with a change of 0.64. The smallest variance was kuiper$\_$V. It changed from 0.11 to 0.05, decreasing by 0.6. Each indicator was changed to 18$\%$ and 11$\%$ when those results were compared to their percentage for the entire parameter distribution.

Redshift dependency was tested with four bins from redshift 0 to 0.5. Most of the indicators were changed by less than 0.1 value for the entire range (see right panel in the Figure \ref{fig:massdepend}). Besides, some of the indicators do not show any prominent trends by redshift changes. The largest variation was shown in stellar mass ratio, which decreased by 0.08, and the smallest variation was demonstrated in kuiper$\_$V, which decreased by 0.04. Kuiper$\_$V showed the least variation by mass and redshift changes; some indicators used the mass show large changes of mean value by mass and redshift variation. 

When we compare the variation by mass and redshift, mass causes more variation than redshift. However, we used clusters having larger than $10^{14} M_{\odot}$, thus dependency is alleviated. It has already been reported that dynamical states dependence on mass of cluster from previous studies \citep{2012MNRAS.427.1322L, 2019ApJ...887..264R, 2021A&A...652A.155S}.

\subsection{Comparison of performance with previous method}
In our previous paper \citep{2024ApJ...970..165K}, we aimed to develop a new linear combination method for dynamical indicators using the rotation matrix. This approach was intended to enhance the separation among different dynamical states. In this study, we compared the performance of our Bayesian classifier with a class-conditional GMM with our earlier method to demonstrate the improvements achieved.

It is important to note that we cannot make direct comparisons due to the different classification and quantification methods employed and the different indicators used. Therefore, we calculated the precision, recall, and accuracy values based on the best linear combination axis in four dynamical indicator spaces of two merger stage classifications in the previous method, using zero as the reference point. These values are presented in Table \ref{tab:compare}. 

The results indicate significant improvements in all metrics (precision, recall, and accuracy) for both merger and relaxed state samples. The Bayesian classifier with GMM for class-conditional distribution method exhibits 20$\%$ to 40$\%$ increases in precision, 6$\%$ to 28$\%$ increases in recall, and 32$\%$ to 41$\%$ increases in accuracy compared to the previous method. Notably, the accuracy shows a remarkable enhancement, yielding twice as significant results as before.

Our previous method provides a continuous probability value for each dynamical state along the x-axis. However, it could not provide specific probabilities for individual clusters. This limitation means that when two clusters share the same x-value but differ in y-value, the method cannot distinguish between them based on the y-value (i.e., the accuracy problem). In contrast, this new classification method in this study can provide specific probabilities for each cluster by considering multidimensional covariances. At this point, the new method could significantly improve the accuracy value.

\begin{table}
	\centering
	\caption{Comparison with our previous study and this study using four indicator combination classification.}
	\label{tab:compare}
\begin{tabular}{l|lll}
\hline
\multicolumn{1}{c|}{} & \multicolumn{1}{c|}{Precision} & \multicolumn{1}{c|}{Recall} & \multicolumn{1}{c}{Accuracy} \\ \hline
\multicolumn{1}{c}{} & \multicolumn{3}{c}{Previous method (Rotation method)} \\ \hline
                      merger & \multicolumn{1}{l|}{0.5} & \multicolumn{1}{l|}{0.6104} &  0.4582 \\ \hline
                      relaxed & \multicolumn{1}{l|}{0.5} & \multicolumn{1}{l|}{0.7217} &  0.5418 \\ \hline
\multicolumn{1}{c}{} &\multicolumn{3}{l}{Current method (GMM)} \\ \hline
                     merger  & \multicolumn{1}{l|}{0.9066} & \multicolumn{1}{l|}{0.8979} & 0.8368 \\ \hline
                     relaxed  & \multicolumn{1}{l|}{0.7671} & \multicolumn{1}{l|}{0.7831} & 0.8368 \\ \hline
\end{tabular}
\end{table}

\subsection{Importance rank of indicators}\label{subsec:bestind}
In the Bayesian classifier with GMM for a class-conditional distribution classification, it is not possible to measure the importance of individual indicators directly. However, we can infer their relative importance based on the best combination results from the comparisons presented in Section \ref{sec:proj}. The combinations generally appear to be organized and show incremental improvements, suggesting that the indicators have significant differences in their contributions.

Despite this initial impression, our testing revealed that the order of the best combinations can change when varying sample sizes, train-test sample ratios, and resampling numbers. One consistently strong indicator is the stellar mass gap, which generally ranks as the best overall. When considering two indicator combinations, the combination of stellar mass gap and center offset is the strongest one, which remains unchanged. 

On the other hand, other indicators seem to have similar importance to each other within the error range and the best combination results change by variable from three to five indicator combinations. This is a consistent result with our previous study, where we suggested that three indicators have comparable importance within the same error range in the different classification methods. However, the analysis methods also affect these results; we discussed it in the appendix \ref{app:1}.

Although the best combinations may vary in three to five indicator combinations depending on variables such as the number of samples, the main probability contours remain relatively unchanged. The region representing 90$\%$ probability stays consistent. This fact demonstrates the superiority and stability of our method.

\subsection{Systematics}\label{sec:Systematics}
In this step, we tested various kinds of systematic errors and their effect on classification results. To mimic the observational systematics error, we tested the indicator distribution with a new cluster galaxy sample by including a different range (1.2, 1.5, 2, 2.5Rvir radius) of line-of-sight interlopers within the three-box samples. In the table \ref{tab:systematics}, the first table shows the percentages of included interlopers. Naturally, when we considered a longer line-of-sight distance, the sample includes more interlopers.

The second table shows classification results by different line-of-sight distances for the interloper sample. When we include more interlopers, fewer consistent classification results are shown, indicating that the fraction of interlopers affects misclassification. This interloper effect is more pronounced during the ancient merger stage. For example, the 1Rvir sample data classified with 45, 22, and 29 clusters for recent merger, ancient merger, and relaxed state. The 1.2Rvir sample classified with 38, 28, and 30 clusters for recent merger, ancient merger, and relaxed state. Among them, 34, 18, and 28 clusters have consistent classification results, and they are about 89, 64, and 93 percent the same as the 1Rvir sample classification. The ancient merger showed the lowest consistency among others. Nevertheless, the consistency of classification is not significantly affected by the interloper fraction and the percentage change is small.

The third table shows the two-sample K-S test p-value changes for each indicator value. For instance, the first low exhibits a two-sample K-S test p-value between the original sample distribution and the 1.2Rvir interloper sample. A small p-value (e.g., < 0.05) indicates that the samples are unlikely to come from the same distribution; most interloper samples differ from the original sample indicator distribution. Although their K-S test p-value of indicator distribution means they are different indicator distributions, their classification results primarily display a good consistency, meaning our classifier also performs consistent classification across different interloper samples. It also means that including interlopers does not significant change the dynamical state classification.

Another possible systematic error is that observed galaxies may suffer confusion when they are projected too close to another galaxy and cannot be separated from the other galaxy. We tested this effect by assigning physical sizes of their stellar disks to the simulated galaxies, following the stellar mass-effective radius relation given in \citet{2019ApJ...880...57M}. We defined confusion as the simulation when a smaller galaxy is totally overlapped by a larger galaxy (the distance between the centers of the two galaxies is smaller than the radius of the larger galaxy minus the radius of the smaller galaxy). From three boxes containing 113 cluster samples, we found that confusion happens very rarely; namely, the average fraction of hidden small galaxies is approximately 0.16$\%$ of the member galaxies. This means that this particular systematic error has a negligible impact on the measurement of dynamical state indicators.

\begin{table}[h!]

\small
\centering
	\caption{Systematics error test results from different line-of-sight range interloper samples.}
	\label{tab:systematics}
\begin{tabular}{|c|c|c|c|}
\hline 
\multicolumn{4}{|c|}{Interloper percentage from the original sample ($\%$)} \\
\hline 
1Rvir & Mean	 & Median	 & Std \\
\hline 
1.2Rvir & 	14 & 	9 & 	13 \\
1.5Rvir & 	19 & 	16 & 	13 \\
2Rvir	 & 23 & 	22 & 	14 \\
2.5Rvir	 & 26 & 	25 & 	14 \\
\hline 
\multicolumn{4}{|c|}{classification result /constant selection number ($\%$)} \\
\hline 
Merger stage  & 	recent & 	ancient & 	relax \\
\hline 
1Rvir & 	45 & 	22 & 	29 \\
1.2Rvir & 	38/34 (89.47) & 	28/18 (64.29) & 	30/28 (93.33) \\
1.5Rvir & 	40/33 (82.50) & 	28/15 (53.57) & 	28/27 (96.43) \\
2Rvir & 	39/32 (82.05) & 	25/14 (56) & 	32/28 (87.50) \\
2.5Rvir & 	39/32 (82.05) & 	27/13 (48.15) & 	30/26 (86.67) \\
\hline 
\multicolumn{4}{|c|}{K-S test p-value of indicators} \\
\hline 
	 & orig-1.2Rvir & 	orig-2.5Rvir & 	1.2Rvir-2.5Rvir \\
     \hline 
$S$ & 	0.895 & 	 0.443 & 	0.796 \\
$\Delta M_{*, 12}$ & 	1.000	 & 0.678 & 	0.993 \\
$d_{off}$ & 	0.443 & 	0.344 & 	0.993 \\
$f_{M_*}$ & 	0.895	 & 0.068 & 	0.344 \\
$V$  & 	0.993 & 	0.962 & 	0.796 \\
$\beta$ & 	0.068 & 	0.002 & 	0.260\\
\hline 
\end{tabular}
\tablefoot{ The first table shows different interloper percentages by the size of the line-of-sight distance. The column shows the mean, median, and standard deviation of interloper percentages. The second table shows the classification result difference. The column shows different merger stage classification results. The third table shows the K-S test p-value of each indicator for different interloper samples. Column shows original -1.2Rvir, original -2.5Rvir, and 1.2Rvir-2.5Rvir line-of-sight interloper sample. See details in section \ref{sec:Systematics}.}
\end{table}

\subsection{Caveats of the method}\label{sec:caveat}
While we demonstrated improved performance in classification and ease of application with the Bayesian classification method, there are several assumptions and caveats to consider. 

First, some indicator calculations are dependent on the depth of data ($\simeq$low magnitude limit) and radial coverage of member galaxies within clusters. Typically, galaxy clusters consist of a small number of massive galaxies and a large number of low-mass galaxies. Mass-based or centrally effective indicators are not significantly affected, but position-based or cluster-wide effective indicators can be heavily influenced by how low-mass galaxies are considered.

Additionally, our method relies on modeling the distribution of indicator values based on simulation data. Although we attempted to eliminate data distribution bias by iteratively resampling the data, our predictions and training model are based on the assumption that our simulation data accurately represents the evolution of galaxy clusters.

Nevertheless, our method can be easily applied to other simulations and observations, and it is also a quick and straightforward process to add more indicators. Spectroscopic or X-ray indicators can be incorporated using hydrodynamical simulation data as long as the data can be adjusted to maintain a consistent depth. Finally, this method can be effectively utilized across various observational datasets.

\section{Conclusions}\label{sec:conclusion}
In this work, we developed an improved method for classifying dynamical states by utilizing a Bayesian classifier with GMM for class-conditional distributions applied to the N-cluster Run simulation data. This Bayesian approach allows us to address the limitations of the previous linear combination method for dynamical state indicators, including restrictions on the number of combined indicators and the challenges posed by non-linear decision boundaries across multiple indicator spaces.

Initially, we classified two merger stages (merger and relaxation) and three merger stages (recent merger, ancient merger, and relaxation) based on the merger mass ratio and time since the merger within the simulation data. Subsequently, we calculated six optical indicators on the projected plane to simulate the observational data.

We modeled each sample of merger stages individually using a GMM. Next, we created cross-conditional distributions of probabilities for both two- and three-merger stages (see Figure \ref{fig:modeling}, Figure \ref{fig:modeling1} and Figure \ref{fig:expsim}). The best configuration was determined through comparisons of precision, recall, and accuracy values. This trained Bayesian classifier was then applied to a reduced set of indicator combinations and observation data (see Figure \ref{fig:dyst}).

Our analysis revealed that varying numbers of Gaussian distributions are required to effectively model each merger stage sample. The performances of the best models are as follows: for the two-merger stage classification, the merger and relaxed sample show average values about 0.92. In the case of the three-merger stage, the recent merger, ancient merger, and the relaxed sample exhibit an average value of about 0.77 (see Table \ref{tab:bestmodel}).
 
When we applied the best classifier derived from six indicator spaces to the lower number of indicator spaces, the results consistently outperformed those of non-projected classifiers. It means that even if we get a small number of indicators, we can classify them with higher performance by utilizing this new method.(See Figure \ref{fig:PRA})

An increase in the number of indicator combinations leads to improved classification results.  (See Figure \ref{fig:PRA}) As a byproduct of the projected classifier analysis, we can get the order of rank for important indicators. Order is as follows: magnitude difference, center offset, sparsity, Kuiper V statistic, and mirror asymmetry. (see Table \ref{tab:bestcombi}) The first two indicators were the same as the result of our previous research.

The new GMM method demonstrates enhanced performance compared to our previous results, which utilized a linear combination of indicators via a rotation matrix. In our comparison of four indicator combinations for two-merger stage classification, the GMM method exhibits improvements ranging from 20$\%$ to 40$\%$ in precision, 6$\%$ to 28$\%$ in recall, and 32$\%$ to 41$\%$ in accuracy (see Table \ref{tab:compare}).

Using this enhanced method and results, we can enlarge our research for the purpose of various future works. In this study, we primarily employed optical indicators; however, we can enhance our methodology by incorporating spectroscopic and X-ray indicators using hydro-magnetic dynamic simulations. Since the inclusion of additional indicators typically improves classification accuracy, we anticipate that the use of spectroscopic indicators will lead to more precise classifications, while X-ray indicators will facilitate a more detailed separation of merger stages.

Furthermore, we aim to create a map of the dynamical states of galaxy clusters by utilizing public photo-z catalogs. Additionally, we will assess the large-scale environmental impact on the dynamical state of these galaxy clusters through this map. This mapping, bolstered by the method and results presented here, will enable us to explore the mass assembly history of the nearby Universe.

The method is available on GitHub\footnote[2]{https://github.com/kimyo1/bcdc}, under a GNU general public license.

\begin{acknowledgements}
Here we thank the anonymous referee for useful comments that have improved this paper. Hectospec observations used in this paper were obtained at the MMT Observatory, a joint facility of the Smithsonian Institution and the University of Arizona. This research was supported by the Agencia Nacional de Investigación y Desarrollo (ANID) ALMA grant funded by the Chilean government, ANID-ALMA-31230016. RS acknowledges financial support from FONDECYT Regular 2023 project No. 1230441 and also gratefully acknowledges financial support from ANID - MILENIO NCN2024$\_$112. M.C. was supported by the EPSRC Prosperity Partnerships grant ARCANE, EP/X025454/1. M. C. and P.T. were supported by the EPSRC Prosperity Partnerships grant ARCANE, EP/X025454/1. P.T. was also supported by the and UKRI Horizon Europe Underwriting, EPSRC EP2293110. YLJ acknowledges support from the Agencia Nacional de Investigaci\'on y Desarrollo (ANID) through Basal project FB210003, FONDECYT Regular projects 1241426 and 123044, Millennium Science Initiative Program NCN2024\_112. HSH acknowledges the support of the National Research Foundation of Korea (NRF) grant funded by the Korea government (MSIT), NRF-2021R1A2C1094577, and Hyunsong Educational \& Cultural Foundation. J.H.S. acknowledges support from the National Research Foundation of Korea grants (No. RS-2025-00516904 and No. RS-2022-NR068800) funded by the Ministry of Science, ICT \& Future Planning. K.W.C. was supported by the National Research Foundation of Korea (NRF) grant funded by the Korea government (MSIT) (2021R1F1A1045622).

\end{acknowledgements}

\bibliographystyle{aa}
\bibliography{example}

\begin{appendix} 

\section{Comparison of best indicator selection by different methods}\label{app:1}

In order to estimate the reliability of the learnt indicator importances, we tested the feature (indicator) importance proposed in section \ref{subsec:bestind} by performing two additional machine learning algorithms, derived within very different frameworks. Generalized Relevance Learning Vector Quantization (GRLVQ, \cite{10.1016/S0893-6080(02)00079-5}) is a prototype-based classification algorithm that compresses dominant properties of the classes into a small set of representative points.

During the learning stage, the relative importance (relevance) of the input features in discriminating between classes (merger stages) is estimated. The framework has been generalized to metric learning, where feature relevances are replaced by full metric tensors, describing the dominant discriminative directions in feature space \cite{schneider_adaptive_2009}.

The restriction of the metric tensor to be diagonal with possibly different diagonal elements collapses to the GRLVQ formulation. While the more general GMLVQ can be considered as a feature construction method, (aggregating features into discriminative directions), its diagonal restriction, GRLVQ, can be interpreted as a feature selection algorithm, because no combination of the input features is performed.

\begin{figure}

\centering
\includegraphics[scale=.62]{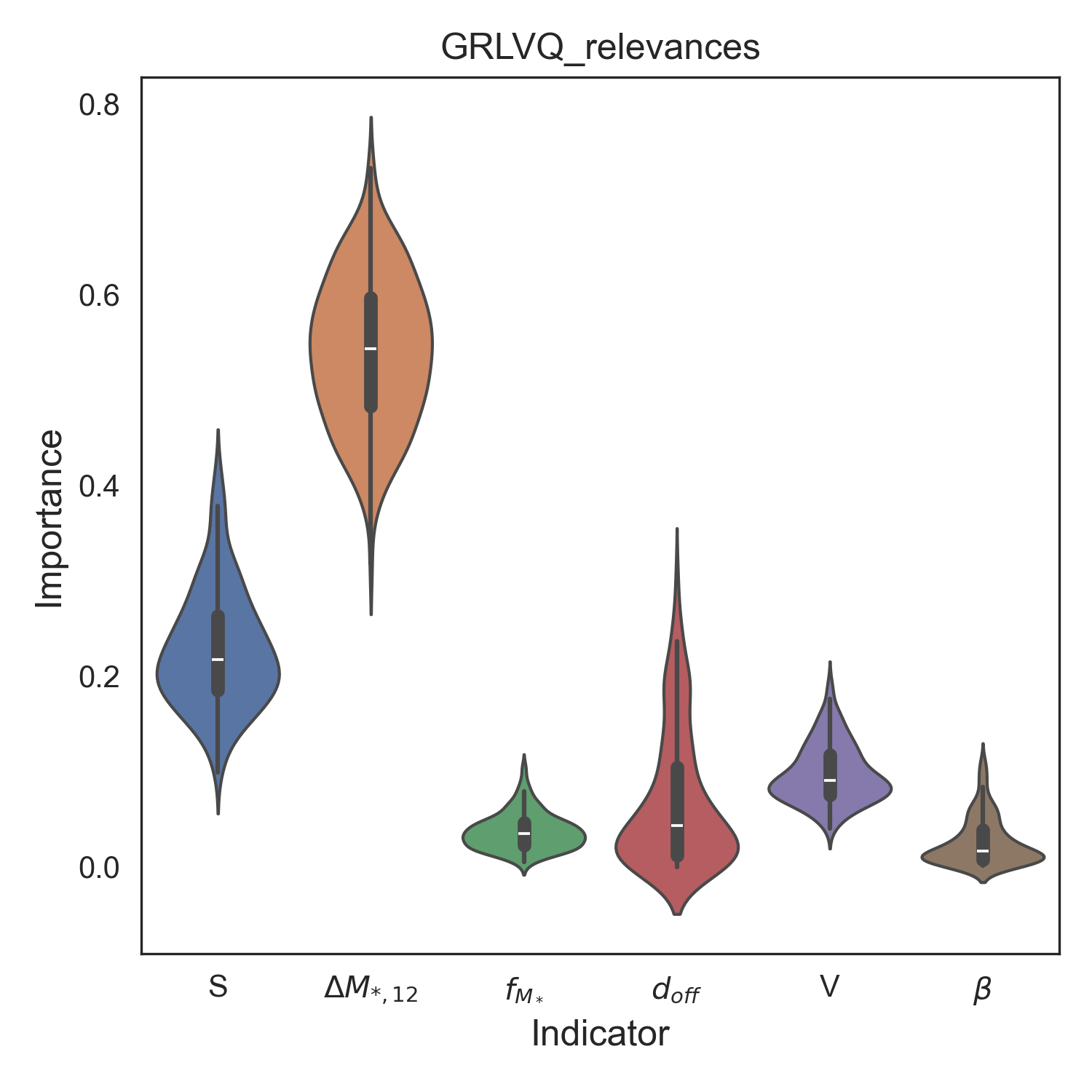} 
\includegraphics[scale=.62]{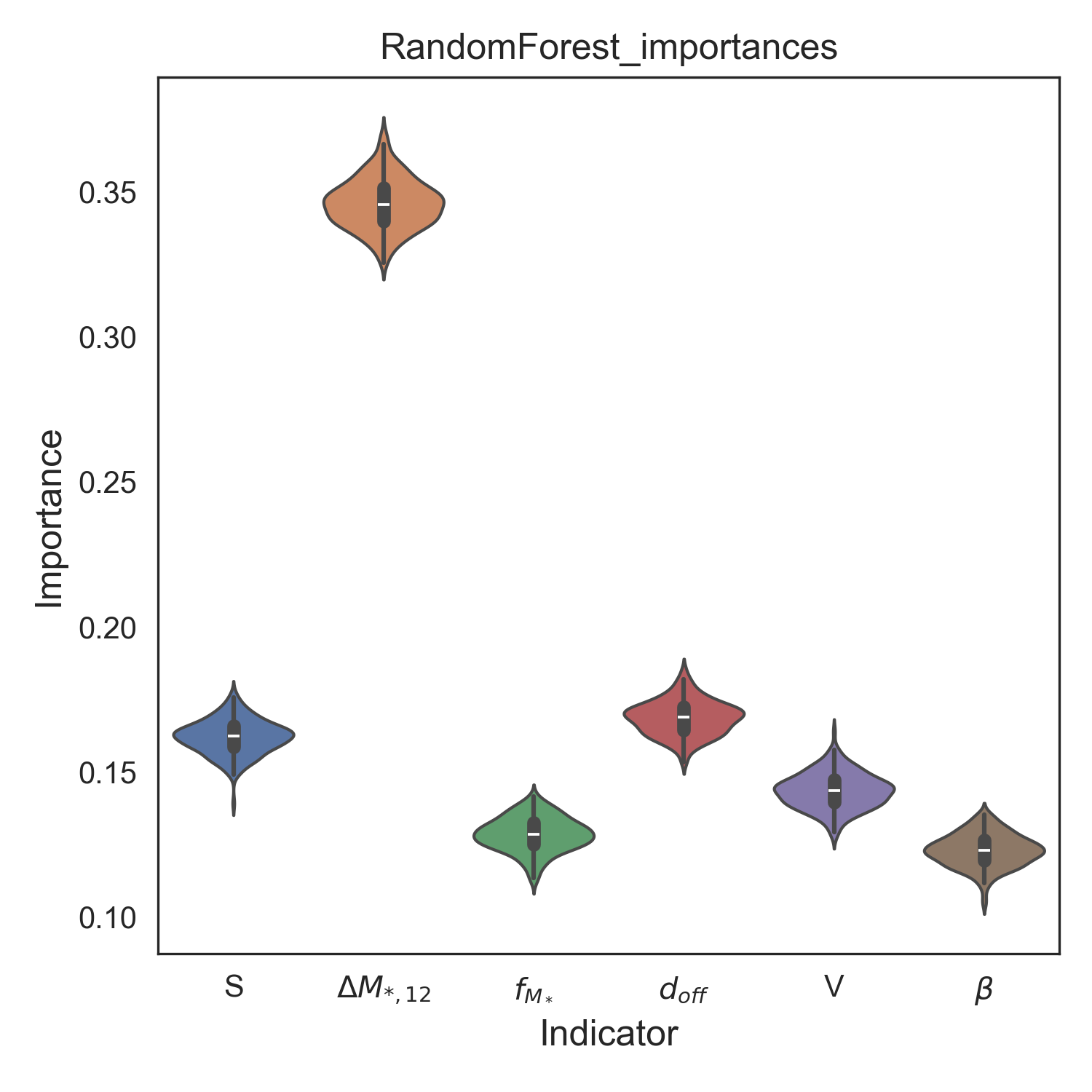}
\caption{Indicator importance test with different methods. Top: Violin plot of indicator relevances from the GRLVQ method. Bottom: Violin plot of indicator importances from the random forest method.  }\label{fig:methodindimp} 
\end{figure}

Given the scope of this work, we opted for GRLVQ, as we would like to compare the feature importances directly with the sequence recovered in section \ref{subsec:bestind}. In this work we adopted the MATLAB implementation available at Michael Biehl's homepage \footnote[3]{https://www.cs.rug.nl/~biehl/gmlvq}, with $5$ prototypes per class, in order to account for non-linear decision boundaries and data sparsity. We perform $500$ independent repetitions of the classification over random sub-samples of the training set and only keep the relevance profiles of the top $40\%$, preserving $200$ models in total. The distribution of feature relevances, across the final $200$ models is shown in Figure \ref{fig:methodindimp}, top panel.

The second algorithm adopted for feature importance determination in a classification setup relies on fitting a random forest \cite{breiman_random_2001} model to the data. A random forest is an ensemble model that agglomerates the performance of multiple decision trees by averaging them. The procedure introduces randomness by constructing multiple decision trees over samples of the data and $/$ or a subset of features. Decision trees tend to overfit on the sample, capturing redundant information disjointly. However, when averaging over the predictions, the independent errors can cancel out, stabilizing the results and providing a lower variance estimator. While by design more opaque than the methodologies presented in this work, RF is a commonly used algorithm for feature selection.

The adopted implementation is the one provided in the \texttt{scikit-learn} package, with permutation importance estimation over $200$ repetitions, bootstrap over training and $100$ base learners (trees). Permutation importance compares the results on any set (we used the full training set) to the model applied on the same set with repeatedly permuted feature. Feature importances are shown in Figure \ref{fig:methodindimp}, bottom panel. It is reassuring to verify that both methodologies recover a similar profile for the feature importances over the training set. These are also comparable with the ones discussed in section \ref{subsec:bestind} up to noise.

In the figures, the importance of the indicators is clearly illustrated by the median value (represented by a white line) within the violin plots. Consistent with the GMM results, the stellar mass gap ($\Delta M_{*,12}$) consistently demonstrates the highest level of importance. However, random forest identifies the center offset ($d_{off}$) as the second most significant indicator, mirroring the findings of the GMM analysis. On the other hand, GRLVQ ranks the sparsity as the second most important indicator. Interestingly, in GRLVQ, the center offset is ranked fourth in importance, which is somewhat unexpected since many other studies have identified it as the second most critical indicator \citep{2019ApJ...887..264R, 2020ApJ...904...36Z, 2020MNRAS.492.6074H}.

It is worth noting that there exist other techniques for feature relevance estimation, and their results provide us with different relevance, although the sample is the same. 

\section{Probabilities of dynamical states for 135 HeCS clusters }\label{app:2}

\begin{table*}[]
\small
\centering
	\caption{Classification results with probabilities for the 135 HeCS clusters.}
	\label{tab:dyst_table1}
\begin{tabular}{|l|l|l|l|l|l|l|l|}
\hline
ID	& R.A. [deg]	& Dec [deg]	& z	& prob$\_$recent& 	prob$\_$ancient& 	prob$\_$relax& 	Classification\\
\hline
A85& 	10.42629& 	-9.4255& 	0.0557& 	0.640037863& 	0.359961014& 	1.12337564E-6& 	recent\\
A119& 	14.03449& 	-1.16356& 	0.0446& 	0.054177855& 	0.746171073& 	0.199651072& 	ancient\\
A160& 	18.25117& 	15.49587& 	0.0432& 	0.518175643	& 0.476778397& 	0.00504595986& 	recent\\
A168& 	18.81422& 	0.26408& 	0.0451& 	0.68586062& 	0.314139092& 	2.88538754E-7& 	recent\\
RXJ0137& 	24.35443& 	-9.27309& 	0.0409& 	0.132576137& 	0.856829319& 	0.0105945442& 	ancient\\
A295& 	30.52788& 	-1.07177& 	0.0424& 	0.219856915& 	0.779028003& 	0.00111508208& 	ancient\\
A602& 	118.50572& 	29.28468& 	0.0606& 	0.672619825& 	0.327380175& 	2.35411865E-26& 	recent\\
Zw1665& 	125.83127& 	4.38181& 	0.0302& 	0.739103548& 	0.0447140134& 	0.216182438& 	recent\\
A671& 	127.13349& 	30.39547& 	0.0503& 	0.500018715& 	0.476520861& 	0.0234604243& 	recent\\
A736& 	136.85728& 	52.11515& 	0.0613& 	0.436803266& 	0.563196203& 	5.30777272E-7& 	ancient\\
A743& 	136.613998& 	10.34499& 	0.1349& 	0.00194692059& 	0.998053079& 	2.58612304E-16& 	ancient\\
A757& 	138.07681& 	47.71318& 	0.0514& 	0.0898227155& 	0.910177285& 	2.5002591E-12& 	ancient\\
A779& 	139.93795& 	33.67961& 	0.0233& 	0.327149627& 	0.672424626& 	4.25747024E-4& 	ancient\\
A957& 	153.42538& 	-0.88957& 	0.0455& 	0.01157835& 	0.05951727& 	0.92890437& 	relax\\
A971& 	154.95695& 	40.98997& 	0.0923& 	0.04345605& 	0.45114644& 	0.50539751& 	relax\\
RXCJ1022.0+3830& 	155.59093& 	38.51348& 	0.0546& 	0.17852275& 	0.727484877& 	0.0939923738& 	ancient\\
A1035B& 	158.0307& 	40.30615& 	0.0801& 	0.75450419& 	0.24549581& 	5.0424738E-10& 	recent\\
A1035A& 	158.12111& 	40.15202& 	0.0675& 	0.12325933& 	0.36992372& 	0.50681695& 	relax\\
A1066& 	159.88703& 	5.17495& 	0.0684& 	0.469852904	& 0.530144523& 	2.57237064E-6& 	ancient\\
A1126& 	163.490295& 	16.850201& 	0.0846& 	0.0744666334& 	0.922689922& 	0.00284344419& 	ancient\\
RXJ1053.7+5450& 	163.53052& 	54.82009& 	0.0727& 	0.356355718& 	0.643639334& 	4.94754996E-6& 	ancient\\
A1142& 	165.23891& 	10.54781& 	0.035& 	0.626442245& 	0.373557755& 	7.50863968E-16& 	recent\\
A1173& 	167.30368& 	41.55554& 	0.076& 	0.30854207& 	0.14908457& 	0.54237336& 	relax\\
A1185& 	167.695007& 	28.705999& 	0.0336& 	0.865605213& 	0.134394772& 	1.50180724E-8& 	recent\\
A1190& 	167.94585& 	40.85659& 	0.0755& 	0.212433489& 	0.787566511& 	5.66097724E-19& 	ancient\\
A1205& 	168.46822& 	2.47489& 	0.0756& 	0.297358848& 	0.702641132& 	1.99117769E-8& 	ancient\\
RXCJ1115.5+5426& 	168.83181& 	54.43409& 	0.0701& 	0.0889304297& 	0.873931866& 	0.0371377043& 	ancient\\
SHK352& 	170.41799& 	2.89252& 	0.0484& 	0.492581864& 	0.506748027& 	6.70108584E-4& 	ancient\\
A1272& 	172.468185& 	23.791901& 	0.1369& 	0.258610265& 	0.69595604& 	0.0454336953& 	ancient\\
A1291B& 	173.03251& 	56.00072& 	0.0582& 	0.380989026& 	0.617303084& 	0.00170789032& 	ancient\\
A1291A& 	173.09313& 	55.97981& 	0.0508& 	0.814620967& 	0.07888229& 	0.106496743& 	recent\\
A1307& 	173.220795& 	14.469& 	0.0811& 	0.871381922& 	0.128618032& 	4.60984194E-8& 	recent\\
A1314& 	173.6175& 	49.08409& 	0.0334& 	0.703567644& 	0.296426337& 	6.01962344E-6& 	recent\\
A1367& 	176.152008& 	19.759001& 	0.0225& 	0.102721718& 	0.897278282& 	4.18227221E-22& 	ancient\\
A1377& 	176.84635& 	55.73779& 	0.0515& 	0.05478425& 	0.20301787& 	0.74219787& 	relax\\
A1424& 	179.38679& 	5.08769& 	0.0754& 	0.100770784& 	0.898467896& 	7.61319862E-4& 	ancient\\
A1436& 	180.0136& 	56.23839& 	0.0648& 	0.253902132& 	0.746097807& 	6.07187946E-8& 	ancient\\
A1446& 	180.499695& 	58.047501& 	0.1027& 	0.711391237& 	0.288608759& 	3.77024888E-9& 	recent\\
MKW4& 	181.11626& 	1.84521& 	0.0204& 	0.110700574& 	0.824745487& 	0.0645539385& 	ancient\\
RXCJ1210.3+0523& 	182.57925& 	5.36913& 	0.0764& 	0.121366864& 	0.680044287& 	0.198588849& 	ancient\\
Zw1215& 	184.4464& 	3.6781& 	0.0772& 	0.01803745& 	0.19080395& 	0.7911586& 	relax\\
NGC4325& 	185.83697& 	10.58173& 	0.0257& 	0.965798606& 	0.0339030735& 	2.98320949E-4& 	recent\\
Virgo& 	186.63401& 	12.72299& 	0.0036& 	0.147261552& 	0.852738448& 	1.0543114E-44& 	ancient\\
A1552& 	187.45088& 	11.80133& 	0.0861& 	0.428855333& 	0.571144667& 	4.71422326E-22& 	ancient\\
A1589& 	190.329605& 	18.571091& 	0.0709& 	0.664434937& 	0.33556493& 	1.3295129E-7& 	recent\\
A1650& 	194.64643& 	-1.57374& 	0.0843& 	0.0309153071& 	0.947218836& 	0.0218658573& 	ancient\\
Coma& 	194.929504& 	27.93862& 	0.0234& 	0.706394967& 	0.282145349& 	0.0114596842& 	recent\\
A1663& 	195.72668& 	-2.50067& 	0.0837& 	0.450082319& 	0.54990531& 	1.2371277E-5& 	ancient\\
A1668& 	195.939789& 	19.27149& 	0.0643& 	0.714405842& 	0.285593557& 	6.01481433E-7& 	recent\\
MS1306& 	197.33011& 	-1.62176& 	0.0832& 	0.511744154& 	0.488148792& 	1.07054716E-4& 	recent\\
A1691& 	197.791107& 	39.217609& 	0.0721& 	0.508701832& 	0.488752284& 	0.00254588358& 	recent\\
A1728& 	200.5719& 	11.2223& 	0.0899& 	0.142226009& 	0.857627907& 	1.46083364E-4& 	ancient\\
RXJ1326.2+0013& 	201.57297& 	0.22139& 	0.0827& 	0.526492704& 	0.342611546& 	0.13089575& 	recent\\
MKW11& 	202.37998& 	11.78861& 	0.0228& 	0.773739453& 	0.226260547& 	1.93913741E-17& 	recent\\
A1750& 	202.70777& 	-1.87455& 	0.0856& 	0.131356919& 	0.868643081& 	4.30725766E-18& 	ancient\\
A1767& 	204.10951& 	59.16231& 	0.0714& 	0.478922256& 	0.520474987& 	6.02756519E-4& 	ancient\\
A1775& 	205.473999& 	26.372& 	0.075& 	0.224349972& 	0.773525399& 	0.00212462911& 	ancient\\
RXJ1342+0213& 	205.522797& 	2.2275& 	0.0773& 	0.504029416& 	0.495675878& 	2.94706101E-4& 	recent\\
A1773& 	205.54655& 	2.21104& 	0.0782& 	0.901262153& 	0.0987378465& 	1.00282057E-9& 	recent\\
A1795& 	207.220703& 	26.5956& 	0.063& 	0.999986325& 	1.36754985E-5& 	1.33067178E-17& 	recent\\
A1800& 	207.340195& 	28.1036& 	0.0755& 	0.03631206& 	0.25981161& 	0.70387632& 	relax\\
RXCJ1351.7+4622& 	208.00093& 	46.23678& 	0.063& 	0.755323635& 	0.244667349& 	9.01630019E-6& 	recent\\
A1809& 	208.3093& 	5.16814& 	0.079& 	0.177711675& 	0.74510367& 	0.0771846557& 	ancient\\
A1831& 	209.823013& 	27.9729& 	0.075& 	0.00217499& 	0.0687389& 	0.92908612& 	relax\\
A1885& 	213.42432& 	43.65979& 	0.0882& 	0.275439969& 	0.525129628& 	0.199430403& 	ancient\\
A1925& 	217.117111& 	56.8829& 	0.1053& 	0.627960276& 	0.371779557& 	2.6016765E-4& 	recent\\
\hline
\end{tabular}
\tablefoot{ Column 1 is the name of the cluster. Column 2, 3, and 4 are right ascension, declination, and redshift. Column 5, 6, and 7 are classified as probabilities of being a recent merger, an ancient merger, and a relaxed state. Column 8 is the classification results based on probabilities.}
\end{table*}
\newpage

\begin{table*}[]
\small
\centering
	\caption{Continued}
	\label{tab:dyst_table2}
\begin{tabular}{|l|l|l|l|l|l|l|l|}
\hline
ID	& R.A. [deg]	& Dec [deg]	& z	& prob$\_$recent& 	prob$\_$ancient& 	prob$\_$relax& 	Classification\\
\hline

MKW8& 	220.16614& 	3.47871& 	0.0271& 	0.136110546& 	0.863785966& 	1.03487573E-4& 	ancient\\
A1986& 	223.2798& 	21.894699& 	0.1171& 	0.316880783& 	0.440791085& 	0.242328132& 	ancient\\
A2018& 	225.277924& 	47.298618& 	0.0872& 	0.585627947& 	0.406873041& 	0.00749901189& 	recent\\
NGC5846& 	227.05705& 	1.63488& 	0.006& 	0.999479295& 	5.20704596E-4& 	1.44187962E-199& 	recent\\
A2029& 	227.729004& 	5.72& 	0.0773& 	0.00245622869& 	0.997543771& 	2.30490863E-11& 	ancient\\
A2051& 	229.141708& 	-0.9484& 	0.1182& 	0.649787447& 	0.350017148& 	1.9540544E-4& 	recent\\
A2064& 	230.25587& 	48.65919& 	0.0738& 	0.026214446& 	0.970645181& 	0.00314037268& 	ancient\\
A2061& 	230.29936& 	30.68344& 	0.0783& 	0.11250645& 	0.487661379& 	0.399832171& 	ancient\\
A2065& 	230.600677& 	27.697411& 	0.0731& 	0.10615141& 	0.35014699& 	0.5437016& 	relax\\
A2063& 	230.7724& 	8.6025& 	0.034& 	0.376908194& 	0.623091806& 	1.13680227E-12& 	ancient\\
A2067& 	230.77915& 	30.86867& 	0.0737& 	0.353303383& 	0.646696605& 	1.17375662E-8& 	ancient\\
A2107& 	234.909988& 	21.789& 	0.0412& 	0.0286956971& 	0.953826807& 	0.0174774962& 	ancient\\
A2110& 	234.93522& 	30.71472& 	0.0971& 	0.00256161& 	0.04162307& 	0.95581532& 	relax\\
RXJ1540+1752& 	235.037994& 	17.878& 	0.0898& 	0.0628375916& 	0.846787609& 	0.0903747997& 	ancient\\
A2124& 	236.22609& 	36.11548& 	0.0677& 	0.12020587& 	0.18492158& 	0.69487255& 	relax\\
A2142& 	239.61176& 	27.1778	& 0.0903& 	0.115096715& 	0.581880853& 	0.303022431& 	ancient\\
A2147& 	240.578003& 	16.02& 	0.0362& 	0.226609149& 	0.773390851& 	3.68857026E-11& 	ancient\\
A2169& 	243.46616& 	49.13285& 	0.0585& 	0.748795553& 	0.251204344& 	1.03403171E-7& 	recent\\
NGC6107& 	244.43291& 	35.05046& 	0.0311& 	0.0950265439& 	0.90421659& 	7.56866606E-4& 	ancient\\
A2199& 	247.16861& 	39.54899& 	0.031& 	0.00111579861& 	0.998855477& 	2.87247817E-5& 	ancient\\
A2197& 	247.47738& 	40.6627& 	0.03& 	0.0396258942& 	0.960374106& 	3.03444977E-14& 	ancient\\
A2245& 	255.6167& 	33.49741& 	0.0868& 	0.279561538& 	0.491027835& 	0.229410627& 	ancient\\
A2244& 	255.62392& 	34.02974& 	0.0997& 	0.479160523& 	0.520772372& 	6.7104917E-5& 	ancient\\
A2249& 	257.453461& 	34.440601& 	0.0849& 	0.785408401& 	0.21459128& 	3.19105226E-7& 	recent\\
A2255& 	258.13949& 	64.04196& 	0.0801& 	0.625892166& 	0.373836964& 	2.70869999E-4& 	recent\\
NGC6338& 	258.84217& 	57.42515& 	0.0286& 	7.79079588E-4& 	0.99922092& 	8.82664687E-179& 	ancient\\
A2399& 	329.36164& 	-7.81858& 	0.0582& 	0.170445486& 	0.825708281& 	0.00384623226& 	ancient\\
RXCJ2214.8+1350& 	333.66181& 	13.82983& 	0.0264& 	0.126760143& 	0.873239833& 	2.33967863E-8& 	ancient\\
A2428& 	334.08476& 	-9.33194& 	0.0836& 	0.00375604& 	0.00731025& 	0.9889337& 	relax\\
A2457& 	338.938324& 	1.475905& 	0.0594& 	0.175505511& 	0.822034572& 	0.00245991723& 	ancient\\
A2593& 	351.10195& 	14.67236& 	0.0415& 	0.105504089& 	0.673077219& 	0.221418692& 	ancient\\
A2670& 	358.57848& 	-10.4166& 	0.0761& 	0.0459604671& 	0.951919888& 	0.00211964476& 	ancient\\
RMJ005105d2p261716d7& 	12.76264& 	26.300369& 	0.2454& 	5.51925443E-5& 	0.999944807& 	1.74241614E-30& 	ancient\\
A267& 	28.176201& 	1.0125& 	0.2291& 	0.33627185& 	0.24504184& 	0.41868631& 	relax\\
A329& 	33.6712& 	-4.5633& 	0.1394& 	0.08234686& 	0.08000516& 	0.83764798& 	relax\\
A383& 	42.014133& 	-3.529228& 	0.1887& 	0.22371417& 	0.03361908& 	0.74266674& 	relax\\
RMJ072729d3p422756d1& 	111.884338& 	42.510319& 	0.1828& 	0.62915067& 	0.370258959& 	5.90371788E-4& 	recent\\
RMJ073720d9p351741d7& 	114.334702& 	35.284691& 	0.2109& 	0.309246481& 	0.685512272& 	0.00524124737& 	ancient\\
RMJ075100d8p173753d8& 	117.815193& 	17.65724& 	0.1863& 	0.0563556954& 	0.923706734& 	0.0199375704& 	ancient\\
RMJ075655d8p383933d2& 	119.263397& 	38.682671& 	0.2172& 	0.238118655& 	0.761881345& 	9.39004237E-17& 	ancient\\
A611& 	120.236748& 	36.056541& 	0.2871& 	0.679111479& 	0.319920699& 	9.6782237E-4& 	recent\\
A646& 	125.546997& 	47.099998& 	0.1273& 	0.708474398& 	0.139301892& 	0.15222371& 	recent\\
RMJ083056d4p322412d2& 	127.69104& 	32.456001& 	0.2551& 	0.227580128& 	0.772382151& 	3.77211127E-5& 	ancient\\
RMJ083513d0p204654d9& 	128.760117& 	20.78112& 	0.177& 	0.833888154& 	0.166111846& 	7.42090537E-38& 	recent\\
A689&  	129.356003& 	14.983& 	0.2789& 	0.985802424& 	0.0141975756& 	1.50913612E-12& 	recent\\
A697& 	130.736206& 	36.362499& 	0.2812& 	0.329570509& 	0.456528968& 	0.213900523& 	ancient\\
A750& 	137.246902& 	11.0444& 	0.164& 	0.967950331& 	0.0320487566& 	9.12581044E-7& 	recent\\
MS0906& 	137.283203& 	10.9925& 	0.1767& 	0.999927604& 	2.39994167E-5& 	4.83968443E-5& 	recent\\
Zw2701& 	148.197998& 	51.890999& 	0.216& 	0.589192669& 	0.410804489& 	2.8421181E-6& 	recent\\
A963& 	154.259995& 	39.048401& 	0.2041& 	0.456273573& 	0.145916589& 	0.397809838& 	recent\\
A980& 	155.627502& 	50.1017& 	0.1555& 	0.524077135& 	0.255863666& 	0.220059199& 	recent\\
Zw3179& 	156.483994& 	12.691& 	0.1422& 	0.975054193& 	0.0101760738& 	0.0147697332& 	recent\\
A1201& 	168.228699& 	13.4448& 	0.1671& 	0.13068455& 	0.12726623& 	0.74204922& 	relax\\
A1204& 	168.332397& 	17.5937& 	0.1706& 	0.871004403& 	0.119056545& 	0.00993905251& 	recent\\
A1423& 	179.341995& 	33.632& 	0.2142& 	0.08861656& 	0.24239748& 	0.66898596& 	relax\\
A1902& 	215.422607& 	37.295799& 	0.1623& 	0.422842908& 	0.577157092& 	1.59232819E-10& 	ancient\\
A1914& 	216.506805& 	37.827099& 	0.166& 	0.444447906& 	0.257638653& 	0.297913441& 	recent\\
RXJ1504& 	226.032104& 	-2.805& 	0.2168& 	0.222182524& 	0.777817352& 	1.24364201E-7& 	ancient\\
A2111& 	234.933701& 	34.4156& 	0.2291& 	0.897566334& 	0.102433666& 	1.07717568E-10& 	recent\\
A2187& 	246.059097& 	41.2383& 	0.1829& 	0.02352511& 	0.01867846& 	0.95779643& 	relax\\
A2219& 	250.089203& 	46.705799& 	0.2257& 	0.153290656& 	0.845507317& 	0.00120202707& 	ancient\\
A2259& 	260.036987& 	27.6702& 	0.1605& 	0.04617962& 	0.00438242& 	0.94943796& 	relax\\
A2261& 	260.612915& 	32.133801& 	0.2242& 	0.151834309& 	0.626112179& 	0.222053513& 	ancient\\
RXJ2129& 	322.41861& 	0.0973& 	0.2339& 	0.06499979& 	0.21888057& 	0.71611964& 	relax\\
A2396& 	328.9198& 	12.5336& 	0.1919& 	0.548332595& 	0.187002887& 	0.264664519& 	recent\\
RMJ220107d7p111805d2& 	330.280731& 	11.29804& 	0.2379& 	0.208944773& 	0.597420249& 	0.193634977& 	ancient\\
A2631& 	354.420593& 	0.276& 	0.2765& 	0.01152513& 	0.00456453& 	0.98391034& 	relax\\
A2645& 	355.320007& 	-9.0275& 	0.2509& 	0.99984829& 	1.51558279E-4& 	1.52101094E-7& 	recent\\
MS2348+2929& 	357.640015& 	29.497999& 	0.1542& 	0.697556381& 	0.302424022& 	1.95966146E-5& 	recent\\
\hline
\end{tabular}
\end{table*}

\end{appendix}
\end{document}